\documentclass[twocolumn,twocolappendix]{aastex701}	
\usepackage[mathlines]{lineno}	

\shorttitle{Image combination for Roman IV}
\shortauthors{K. Cao et al.}
\usepackage{CJKutf8}

\usepackage{graphicx}	
\usepackage{amsmath}	
\usepackage{amssymb}	
\usepackage{multirow}

\definecolor{F184}{rgb}{0.65,0.1,0.0}
\definecolor{H158}{rgb}{0.35,0.4,0.0}
\definecolor{J129}{rgb}{0.0,0.4,0.35}
\definecolor{Y106}{rgb}{0.0,0.1,0.65}

\newcommand{\papone}{Paper I}
\newcommand{\paptwo}{Paper II}
\newcommand{\papthree}{Paper III}

\begin{document}

\title{Simulating Image Coaddition with the Nancy Grace Roman Space Telescope. IV. Hyperparameter Optimization and Experimental Features}

\author[orcid=0000-0002-1699-6944]{Kaili Cao (\begin{CJK*}{UTF8}{gbsn}曹开力\end{CJK*})}
\affiliation{Center for Cosmology and AstroParticle Physics (CCAPP), The Ohio State University, 191 West Woodruff Ave, Columbus, OH 43210, USA}
\affiliation{Department of Physics, The Ohio State University, 191 West Woodruff Ave, Columbus, OH 43210, USA}
\email[show]{cao.1191@osu.edu}

\author[orcid=0000-0002-2951-4932]{Christopher M. Hirata}
\affiliation{Center for Cosmology and AstroParticle Physics (CCAPP), The Ohio State University, 191 West Woodruff Ave, Columbus, OH 43210, USA}
\affiliation{Department of Physics, The Ohio State University, 191 West Woodruff Ave, Columbus, OH 43210, USA}
\affiliation{Department of Astronomy, The Ohio State University, 140 West 18th Avenue, Columbus, OH 43210, USA}
\email{hirata.10@osu.edu}

\author[orcid=0000-0002-6111-6061]{Katherine Laliotis}
\affiliation{Center for Cosmology and AstroParticle Physics (CCAPP), The Ohio State University, 191 West Woodruff Ave, Columbus, OH 43210, USA}
\affiliation{Department of Physics, The Ohio State University, 191 West Woodruff Ave, Columbus, OH 43210, USA}
\email{laliotis.2@osu.edu}

\author[orcid=0000-0003-1585-997X]{Masaya Yamamoto}
\affiliation{Department of Physics, Duke University, Box 90305, Durham, NC 27708, USA}
\affiliation{Department of Astrophysical Sciences, Princeton University, Princeton, NJ 08544, USA}
\email{my5871@princeton.edu}

\author[orcid=0009-0002-7190-9775]{Emily Macbeth}
\affiliation{Center for Cosmology and AstroParticle Physics (CCAPP), The Ohio State University, 191 West Woodruff Ave, Columbus, OH 43210, USA}
\affiliation{Department of Physics, The Ohio State University, 191 West Woodruff Ave, Columbus, OH 43210, USA}
\affiliation{Department of Astronomy, The Ohio State University, 140 West 18th Avenue, Columbus, OH 43210, USA}
\affiliation{Department of Astronomy/Steward Observatory, University of Arizona, 933 North Cherry Avenue, Tucson, AZ 85721, USA}
\email{macbeth.12@osu.edu}

\author[orcid=0000-0002-5622-5212]{M.~A.~Troxel}
\affiliation{Department of Physics, Duke University, Box 90305, Durham, NC 27708, USA}
\email{michael.troxel@duke.edu}

\collaboration{all}{Roman HLIS Cosmology PIT}

\begin{abstract}

For weak gravitational lensing cosmology with the forthcoming Nancy Grace Roman Space Telescope, image coaddition, or construction of oversampled images from undersampled ones, is a critical step in the image processing pipeline. In the previous papers in this series, we have re-implemented the {\sc Imcom} algorithm, which offers control over point spread functions in coadded images, and applied it to state-of-the-art image simulations for Roman. In this work, we systematically investigate the impact of {\sc Imcom} hyperparameters on the quality of measurement results. We re-coadd the same $16$ blocks ($1.75 \times 1.75 \,{\rm arcmin}^2$, $2688 \times 2688$ pixels each) from OpenUniverse2024 simulations with $26$ different configurations in each of $5$ bands. We then compare the results in terms of $12$ objective evaluation criteria, including internal diagnostics of {\sc Imcom}, properties of coadded noise frames, measurements of injected point sources, and time consumption. We demonstrate that: i) the Cholesky kernel is the best known linear algebra strategy for {\sc Imcom}, ii) for our measurements, a wide Gaussian target output PSF outperforms a smoothed Airy disk or a narrow Gaussian, iii) kernel-specific settings are worth considering for future coaddition, and iv) {\sc Imcom} experimental features studied in this work are either inconsequential or detrimental. We end this paper by discussing current and next steps of {\sc Imcom}-related studies in the context of Roman shear and clustering measurements.

\end{abstract}

\keywords{Astronomy image processing (2306) --- Weak gravitational lensing (1797)}	

\section{Introduction} \label{sec:intro}

Weak gravitational lensing is a promising but demanding cosmological probe. Being directly sensitive to the mass distribution in the Universe, it is free from biases introduced when cosmologists use luminous objects as tracers. Therefore, it contains valuable information about the growth of cosmic structure \citep[e.g.,][]{2001PhR...340..291B, 2013PhR...530...87W, 2015RPPh...78h6901K}. Meanwhile, weak lensing cosmology relies on high-precision measurements of galaxy shapes \citep[see][for a recent review]{2018ARA&A..56..393M}. In the weak lensing scenario, gravitation only causes shape distortions at the percent level, more than an order of magnitude smaller than intrinsic shapes of galaxies. Consequently, weak lensing signals can only be studied statistically, and reliable image processing is a prerequisite for realizing its potential.

As we step into the second half of the 2020s, we are exhilarated to see the successful completion of Stage III surveys and the inauguration of Stage IV missions. Three large programs of the past decade --- the Dark Energy Survey \citep{2022PhRvD.105b3514A, 2022PhRvD.105b3515S}, the Hyper Suprime Cam \citep{2019PASJ...71...43H, 2020PASJ...72...16H, 2023PhRvD.108l3518L, 2023PhRvD.108l3519D}, and the Kilo Degree Survey \citep{2022A&A...664A.170V, 2023A&A...679A.133L, 2025arXiv250319441W} --- have yielded few percent level constraints on cosmological parameters. The ongoing and upcoming surveys --- the Euclid space telescope \citep{2011arXiv1110.3193L, 2022A&A...662A.112E, 2024arXiv240513491E} launched two years ago, the Legacy Survey of Space and Time at the NSF-DOE Vera C. Rubin Observatory (hereafter ``Rubin;'' \citealt{2012arXiv1211.0310L, 2019ApJ...873..111I}), which saw its first light earlier this year, and the Nancy Grace Roman Space Telescope (hereafter Roman; \citealt{2019arXiv190205569A}), which is on track to be launched next year --- are expected to push the precision to the sub-percent level.

After launch, Roman is planned to start a five-year mission at Sun-Earth Lagrange Point 2 (L2). With its Hubble-sized primary mirror ($2.4 \,{\rm m}$) and native pixels ($0.11 \,{\rm arcsec}$), large field of view ($0.281 \,{\rm deg}^2$, over $300$ million active pixels), and high-sensitivity H4RG-10 detectors \citep{2020JATIS...6d6001M}, Roman is a versatile telescope suitable for multiple surveys in infrared. Specifically, its weak lensing program will be implemented with its High Latitude Wide Area Survey (HLWAS). Roman will cover $2400 \,{\rm deg}^2$ of the sky in three bands (``Medium Tier'') and additional $2700 \,{\rm deg}^2$ in the H158 band (``Wide Tier''), yielding unprecedented galaxy number densities of $41.3 \,{\rm arcmin}^{-2}$ and $26.7 \,{\rm arcmin}^{-2}$, respectively \citep{2025arXiv250510574O}.

As a space telescope, the Roman point spread function (PSF) is not limited by seeing conditions of the Earth's atmosphere, and will be both more stable and narrower than those of ground-based instruments operating in similar bands. While high resolution is desirable, narrow PSFs also create a challenge: To enable an efficient survey, the native pixel size of Roman is larger than what is needed to fully resolve its diffraction-limited PSFs ($\lesssim \lambda/2D$, where $\lambda$ is the wavelength of observation and $D$ is the entrance pupil diameter), yet this full resolution is required for breaking degeneracy of Fourier modes and enabling accurate shape measurements. To meet this challenge, Roman will dither its camera and take several undersampled images of the same area of the sky, which we will then combine to construct oversampled images.

Such combination is usually referred to as image coaddition and formulated as a linear transformation from input pixels to output pixels \citep[see][for the necessity of linearity]{2023OJAp....6E...5M}. Traditional algorithms like {\sc Drizzle} \citep{2002PASP..114..144F, 2012drzp.book.....G} assign coaddition weights by computing geometric overlaps between input and output pixels. This is efficient, but the resulting output images lack well-defined PSFs, and how to calibrate weak lensing shear estimators is unclear. By building and solving linear systems, the {\sc Imcom} technique \citep{2011ApJ...741...46R} minimizes discrepancies between as-realized output PSFs and user-specified target PSFs, and thus provides coadded images with uniform PSFs. Furthermore, it can handle arbitrary rolls, distortions, missing pixels, and dithering patterns, which is useful for addressing real-world issues in actual surveys.

This series of papers has been focused on applying {\sc Imcom} to Roman image processing. \citet[][hereafter \papone]{2024MNRAS.528.2533H} re-implemented {\sc Imcom} as a Python program with a C back end (the original implementation by \citealt{2011ApJ...741...46R} is in Fortran), enabled coaddition of larger areas of the sky using a divide-and-conquer strategy, and tested it using Roman-like images simulated by \citet{2023MNRAS.522.2801T}. \citet[][hereafter \paptwo]{2024MNRAS.528.6680Y} further diagnosed the output images in terms of noise properties of coaddition results and measurements of simulated point sources; systematic errors introduced by {\sc Imcom} were found to meet Roman requirements. \citet[][hereafter \papthree]{2025ApJS..277...55C} reorganized the program into an object-oriented framework known as {\sc PyImcom}, employed various measures to make it more efficient, and introduced new linear algebra strategies (referred to as ``kernels'') for determining coaddition weights. {\sc PyImcom} was used for coadding state-of-the-art OpenUniverse2024 \citep[][hereafter OU24]{2025arXiv250105632O} simulated images.

The {\sc PyImcom} configuration interface allows users to specify a myriad of hyperparameters (i.e., parameters that are determined before running the program), including but not limited to target output PSFs. These hyperparameters were empirically configured for previous simulations; in this work, we systematically investigate their impact on the quality of output images. This paper is structured as follows. In Section~\ref{sec:meth}, we review the {\sc Imcom} formalism, introduce new features accompanying OU24 images, and define objective evaluation criteria for {\sc Imcom} outputs. In Section~\ref{sec:config}, we detail the configuration of tests conducted in this work. For both Cholesky and iterative kernels, we explore the choice of target output PSFs, kernel-specific settings, and some experimental features. In Section~\ref{sec:base}, we present benchmark results using both kernels in five bands. Then in Section~\ref{sec:var}, we compare variant cases to benchmark cases and each other. Two tables supplementing visualizations in these two sections are included in Appendix~\ref{app:tables}. Finally, in Section~\ref{sec:disc}, we conclude this paper by recapping the key results and discussing ongoing and planned {\sc Imcom}-related studies.

\section{{\sc Imcom} Methodology} \label{sec:meth}

This section reviews the status quo of the {\sc Imcom} methodology. In Section~\ref{ss:recap}, we recap the {\sc Imcom} formalism \citep{2011ApJ...741...46R} and its {\sc PyImcom} implementation (\papthree). Then in Section~\ref{ss:ou24}, we present the OU24 image simulations \citep{2025arXiv250105632O} and some new {\sc Imcom} features accompanying them. We describe the $12$ criteria for evaluating {\sc Imcom} results in Section~\ref{ss:eval}.

\subsection{Recap of {\sc Imcom} and {\sc PyImcom}} \label{ss:recap}

We briefly recap the aspects of {\sc Imcom} relevant to the analysis and interpretation of its outputs; the reader is referred to \citet{2011ApJ...741...46R} for the mathematical formalism, \papone\ for the problem statement in the Roman context, and \papthree\ for full details of the {\sc PyImcom} implementation.

Let us consider a set of $n$ input pixels (indexed by Latin letters, e.g., $i=0... n-1$) and $m$ output pixels (indexed by Greek letters, e.g., $\alpha=0... m-1$).\footnote{Since the implementation is in Python, we follow the Python indexing scheme in this paper, and start arrays with 0.} Images are two-dimensional arrays of pixels, but they are flattened here to formulate linear systems; furthermore, pixels from different input images are concatenated into a single vector. In the context of coadding Roman images, roughly speaking, we have $n \sim {\rm several} \times 10^3$ and $m \sim {\rm a\ few} \times 10^3$ for each postage stamp. (In {\sc Imcom}, we divide the sky into postage stamps of size $\sim 1 \,{\rm arcsec}$ to keep linear systems manageable, and tile them to obtain output images for larger areas of the sky.)

A linear image coaddition algorithm attempts to construct an output image $H_\alpha$ from input images $I_i$ with coaddition weights $T_{\alpha i}$:
\begin{equation}
    H_\alpha = \sum_{i=0}^{n-1} T_{\alpha i} I_i.
    \label{eq:coadd}
\end{equation}
For each output pixel $\alpha$, such a linear transformation also constructs a coadded PSF, consisting of the appropriately translated input PSFs:
\begin{equation}
    {\rm PSF}_{\alpha,\rm out}({\boldsymbol R}_\alpha -{\boldsymbol s})
    = \sum_{i=0}^{n-1} T_{\alpha i} G_i({\boldsymbol r}_i - {\boldsymbol s}),
    \label{eq:outpsf}
\end{equation}
where ${\boldsymbol R}_\alpha$ is the position of output pixel $\alpha$, ${\boldsymbol r}_i$ is that of input pixel $i$, and $G_i$ denotes the PSF at ${\boldsymbol r}_i$ in the image containing pixel $i$. {\sc Imcom} attempts to find optimal coaddition weights $T_{\alpha i}$ that minimize
\begin{equation}
    U_\alpha = \left\Vert {\rm PSF}_{\alpha,\rm out} - \Gamma \right\Vert^2
    ~~{\rm and}~~
    \Sigma_\alpha = \sum_{i,j} N_{ij} T_{\alpha i}T_{\alpha j},
    \label{eq:U_Sigma}
\end{equation}
where $\Gamma$ is a uniform ``target'' PSF specified by the user (see Section~\ref{ss:outpsf-m} for common choices), $\Vert\cdot\Vert$ represents the $L^2$ norm, and $N_{ij}$ is the input noise covariance. We refer to $U_\alpha$ as a ``PSF leakage'' metric and $\Sigma_\alpha$ as a ``noise amplification'' metric.

In {\sc Imcom}, we usually assume that the input noise covariance is the identity matrix, i.e., $N_{ij} = \delta_{ij}$ (the Kronecker delta). In other words, we assume that input noise is uniform and uncorrelated. Note that this assumption only applies to the optimization of coaddition weights; any non-identity noise covariance \citep[as it will be for Roman; e.g.,][]{2024PASP..136l4506L} propagates to the output noise covariance (of which $\Sigma_\alpha$ is the diagonal) via Equation~(\ref{eq:coadd}) and can be studied using simulated noise fields (see below). In principle, it is possible to use any $N_{ij}$ in Equation~(\ref{eq:U_Sigma}); nevertheless, the full noise covariance of a Roman sensor chip assembly (SCA) would be a $4088^2 \times 4088^2$ matrix, which is impractical to handle. We thus leave implementation of non-identity input noise covariance for future work if it is shown to be essential. Under the assumption of $N_{ij} = \delta_{ij}$, we have $\Sigma_\alpha = \sum_{i} T_{\alpha i}^2$.

For each output pixel $\alpha$, {\sc Imcom} attempts to minimize a linear combination of PSF leakage and noise amplification, $U_\alpha + \kappa_\alpha \Sigma_\alpha$, where $\kappa_\alpha$ is a Lagrange multiplier. Note that $\kappa_\alpha$ balances two optimization goals, small $U_\alpha$ and small $\Sigma_\alpha$. Following \citet{2011ApJ...741...46R}, \papone\ determined $\kappa_\alpha$ using a bisection search; in \papthree, we found that pre-setting a uniform value for all output pixels is both reasonable and efficient (see Section~\ref{ss:kernel-m} for specific values). Regardless of how $\kappa_\alpha$ is determined, for a given $\kappa_\alpha$, the optimal coaddition weights are
\begin{equation}
    T_{\alpha i} =\sum_j [({\mathbf A} + \kappa_\alpha {\mathbb I}_{n\times n})^{-1}]_{ij} \left(-\frac12B_{\alpha j}\right),
    \label{eq:T-AB}
\end{equation}
where system matrices ${\mathbf A}$ and ${\mathbf B}$ capture PSF overlaps between all pairs of input pixels and between all input pixels and the output pixel, respectively. We refer the reader to \citet{2011ApJ...741...46R} for their respective definitions and to \papthree\ for how {\sc PyImcom} manages them.

As the matrix inverse symbol $^{-1}$ in Equation~(\ref{eq:T-AB}) indicates, {\sc Imcom} needs to solve linear systems. In \papthree, we introduced several linear algebra strategies (kernels); here we focus on the two that are tested in this work, the Cholesky kernel and the iterative kernel. For a small number of $\kappa_\alpha$ values, we can avoid expensive eigendecomposition and use Cholesky decomposition instead. The Cholesky kernel is efficient and has good control over PSF leakage, but is subject to postage stamp boundary effects due to its selection of input pixels (which is shared among all output pixels within each postage stamp; see Figure~1 of \papthree). The iterative kernel selects input pixels in a more symmetric way, builds different linear systems for indivual output pixels, and solves them using the conjugate gradient method \citep{hestenes1952methods}. It has better control over noise, but due to its finite tolerance, it is not as accurate as the Cholesky kernel in terms of PSF construction. Despite their differences, both kernels fall into the category of linear image coaddition and can be compared on a common basis.

Once coaddition weights are computed, {\sc Imcom} can construct output images using Equation~(\ref{eq:coadd}) and calculate internal diagnostics $U_\alpha$ and $\Sigma_\alpha$ using Equation~(\ref{eq:U_Sigma}). An important feature of linear image coaddition Equation~(\ref{eq:coadd}) is that the coaddition weights ($T_{\alpha i}$) only depend on input ($G_i$) and target ($\Gamma$) PSFs, not input signals ($I_i$). This allows us to coadd multiple versions of the images, referred to as layers, using the same set of coaddition weights. While there will only be one version in the real mission, it is useful to accompany actual images with noise fields and injected sources for testing purposes. Specifically, this work makes use of four layers:
\begin{itemize}
    \item {\tt \textquotesingle SCI\textquotesingle}: simulated science images from OU24 \citep{2025arXiv250105632O}. Like in \papthree, this is used for visual validation, while quantitative analyses are based on the following three layers.
    \item {\tt \textquotesingle whitenoise10\textquotesingle}: simulated white (i.e., uncorrelated) noise frames, implemented as Gaussian random fields with mean $0$ and variance $1$.
    \item {\tt \textquotesingle 1fnoise9\textquotesingle}: simulated $1/f$ (correlated in a specific way) noise frames, implemented as an scale-invariant array with unit variance per logarithmic range in frequency for each readout channel.
    \item {\tt \textquotesingle gsstar14\textquotesingle}: injected stars drawn by {\sc GalSim} \citep{Rowe2015A&C}, implemented as ideal point sources located at HEALPix nodes with ${\tt NSIDE} = 14$.
\end{itemize}
We refer readers to Section~3 of \papone\ for further details about noise realizations and other layers coadded in \papone.

\subsection{Coadding OpenUniverse2024 Images} \label{ss:ou24}

Both \papone\ and \papthree\ coadded simulated images from \citet{2023MNRAS.522.2801T}. At the end of 2023, a new suite of simulations were run at the Theta supercomputer at the US Department of Energy's (DOE) Argonne National Laboratory right before its retirement. This simulation suite, known as OpenUniverse2024 \citep{2025arXiv250105632O}, is a joint effort among multiple Rubin and Roman collaborations. It produced simulated images for $\sim 70 \,{\rm deg}^2$ of the Rubin Wide-Fast-Deep survey and the Roman HLWAS, as well as overlapping versions of the Rubin ELAIS-S1 Deep-Drilling Field and the Roman High-Latitude Time-Domain Survey (HLTDS). For a fuller discussion of OU24 data, tools, and features, see \citet{2025arXiv250105632O}. Comparing the Roman arms of both simulation suites, it is important to note that, in addition to the Y106, J129, H158, and F184 bands included in \citet{2023MNRAS.522.2801T}, OU24 also included the redder K213 band and the wide W146 band.

\begin{table*}[]
    \centering
    \caption{\label{tab:base_config}Sizes and dimensions used in this work. The first three and last columns are from Table~3 of \papone; they are included here for clarification and comparison purposes. The fourth and fifth columns present values used for the Cholesky and Iterative linear algebra kernels, respectively. We refer readers to Figure~4 of \papone\ for a diagram of these quantities.}
    \begin{tabular}{llcccc}
    \hline
    Parameter or variable name & Description & \papone & Cholesky & Iterative & Unit \\
    \hline
    {\tt s\_in} & Input (native) pixel scale & $0.11$ & $0.11$ & $0.11$ & ${\rm arcsec}$ \\
    $\Delta\theta$ ({\tt d\_theta}) & Output pixel scale & $0.025$ & $0.0390625$ & $0.0390625$ & ${\rm arcsec}$ \\
    $n_2$ & Postage stamp size in output pixels & $50$ & $32$ & $32$ & \\
    $k$ ({\tt fade\_kernel}) & Transition width in pixels & $3$ & $3$ & $0$ & \\
    $n_2 \Delta\theta$ & Postage stamp angular size (excluding transition region) & $1.25$ & $1.25$ & $1.25$ & ${\rm arcsec}$ \\
    $(n_2+2k) \Delta\theta$ & Postage stamp angular size (including transition region) & $1.4$ & $1.484375$ & $1.25$ & ${\rm arcsec}$ \\
    {\tt INPAD} & Acceptance radius for input pixels & $1.25$ & $1.24$ & $0.6$ & ${\rm arcsec}$ \\
    $n_1$ & Block size in postage stamps (1D) & $48$ & $80$ & $80$ & \\
    {\tt PAD} & Padding region of block in postage stamps & $2$ & $2$ & $2$ & \\
    $n_1 n_2 \Delta\theta$ & Block angular size (excluding extra postage stamps) & $1.0$ & $1.66667$ & $1.66667$ & ${\rm arcmin}$ \\
    $(n_1+2{\tt PAD}) n_2 \Delta\theta$ & Block angular size (including extra postage stamps) & $1.08333$ & $1.75$ & $1.75$ & ${\rm arcmin}$ \\
    $(n_1+2{\tt PAD}) n_2$ & Block image side length in output pixels & $2600$ & $2688$ & $2688$ & \\
    {\tt BLOCK} & Mosaic size in blocks (1D) & $48$ & $36$ & $36$ & \\
    ${\tt BLOCK} \times n_1 n_2 \Delta\theta$ & Mosaic angular size & $0.8$ & $1.0$ & $1.0$ & ${\rm degree}$ \\
    \hline
    \end{tabular}
\end{table*}

OU24 data products included a $1.0 \times 1.0\,{\rm deg}^2$ mosaic in five bands of the simulated Roman HLWAS processed by {\sc PyImcom}; due to difficulties with chromatic PSFs \citep[e.g.,][]{2025MNRAS.542..608B}, W146 images have not been processed by {\sc Imcom}. Based on our experience from \papone\ and \paptwo, as well as the differences between the two simulation suites (most importantly, the inclusion of charge diffusion in OU24, which widens input PSFs; see Section~5.4 of \papone\ for discussion), we have chosen different parameters for {\sc Imcom} coadds. In this work, we adopt those settings\footnote{\url{https://github.com/Roman-HLIS-Cosmology-PIT/pyimcom/tree/main/configs/production_configs_spring2024}} for the benchmark case of the Cholesky kernel, and adjust some of them for the iterative kernel. In Table~\ref{tab:base_config}, we compare sizes and dimensions used in \papone\ (third column), for OU24 coadds and the Cholesky kernel (fourth column), and for the iterative kernel (fifth column); some other settings are discussed in Section~\ref{sec:config}. Basically, the output pixel scale $\Delta\theta$ was enlarged from $0.025 \,{\rm arcsec}$ to $0.0390625 \,{\rm arcsec}$ (by a factor of $56.25\%$),\footnote{For a native PSF in the Y106 band to be Nyquist sampled, the maximum pixel scale is $\sim 0.044869 \,{\rm arcsec}$.} and the parameters for the number pixels ($n_1$, $n_2$, and ${\tt BLOCK}$) were adjusted accordingly, resulting in a mosaic with a larger angular size but less pixels. As explained in \papthree, the iterative kernel does not need transition regions between postage stamps, but requires a smaller acceptance radius, hence we choose different $k$ and ${\tt INPAD}$ for it; neither of these affect the configuration of the output pixel grid.

Two other {\sc PyImcom} features developed to accompany OU24 simulated images are worth emphasizing here. First, \citet{2023MNRAS.522.2801T} only made one PSF at the center of each image (``input'' from the perspective of {\sc Imcom}), and we assumed that it was a constant function in the input pixel plane. (It was not a constant in the output pixel plane because of differences between input and output world coordinate systems.) OU24 made four PSFs at the corners of each image, and {\sc PyImcom} uses Legendre polynomials to handle this spatial variation of input PSFs. Second, in addition to the layers mentioned in Section~\ref{ss:recap}, OU24 coadds included several newly developed ones for various purposes:
\begin{itemize}
    \item {\tt \textquotesingle nstar14,2e5,86,3\textquotesingle}: noisy stars, normalized to total flux of $2 \times 10^5$ electrons with self-Poisson noise, including $86$ $e^2$/input pixel background variance.
    \item {\tt \textquotesingle gstrstar14\textquotesingle}: like {\tt \textquotesingle gsstar14\textquotesingle}, but on in only one of the passes of HLWAS (to test transient response).
    \item {\tt \textquotesingle gsfdstar14,0.05\textquotesingle}: stars with total flux that varies by $5\%$ from center to edge of the focal plane (to test what happens when the filter bandpass varies; $5\%$ is highly exaggerated).
    \item {\tt \textquotesingle gsext14,seed=*,shear=*:*\textquotesingle}: {\sc GalSim} extended objects, exponential profiles with random (with specified seed) orientations, sheared by specified amounts in $g_1$ and $g_2$ directions.
\end{itemize}
We are currently working with other members of the Roman HLIS Cosmology PIT on the analysis of these layers, which are not included in the simulations for this work.

\subsection{{\sc Imcom} Evaluation Criteria} \label{ss:eval}

While a visual inspection is usually an informative first step after getting {\sc Imcom} coadds, we need to define a set of qualitative and objective evaluation criteria for comparison purposes. The {\sc Imcom} formalism contains two internal diagnostics, and we introduced two more in \papthree. In \paptwo, we analyzed {\sc Imcom} outputs in terms of simulated noise frames and simulated (both realistic and ideal) point sources. Besides, time consumption is also an important factor when choosing how to configure {\sc Imcom}. In this section, we present the $12$ {\sc Imcom} evaluation criteria used throughout the rest of this paper.

The first set of four criteria correspond to internal diagnostics of {\sc Imcom}:
\begin{enumerate}
    \item PSF leakage $U_\alpha$, defined in Equation~(\ref{eq:U_Sigma});
    \item Noise amplification $\Sigma_\alpha$, defined in Equation~(\ref{eq:U_Sigma});
    \item Total input weight $T_{{\rm tot}, \alpha}$, defined as a summation of coaddition weights over all input pixels:
    \begin{equation}
        T_{{\rm tot}, \alpha} = \sum_i T_{\alpha i}.
        \label{eq:Tsum}
    \end{equation}
    For each output pixel, this is expected to be close to $1$ (since {\sc Imcom} tries to conserve surface brightness) but not exactly $1$ (due to differences between input and output PSFs); see Section~2.4 of \citet{2025arXiv251016110C} for an intuitive explanation. The uniformity of total input weights among output pixels is desirable.
    \item Effective coverage $\bar{n}_{{\rm eff}, \alpha}$, designed to assess the distribution of contributions from different input images:
    \begin{equation}
        \bar{n}_{{\rm eff}, \alpha} = \frac{
        \left( \sum_{\bar i} |t_{\alpha\bar i}| \right)^2
        }{ \sum_{\bar i} t_{\alpha\bar i}^2},
        ~~~~ t_{\alpha\bar i} = \sum_{i\in\bar{i}} T_{\alpha i},
        \label{eq:Neff}
    \end{equation}
    where $\bar{i}$ denotes the set of input pixel indices corresponding to each input image. This quantity is normalized so that its maximum is the number of input images that geometrically overlap with the output pixel. It also accounts for masked input pixels and assignment of coaddition weights.
\end{enumerate}
Since each of these diagnostics has one value per output pixel, and there are millions of pixels in each block, some downsampling is necessary. Following \papthree, we study either the average (most of them) or the standard deviation (total input weight only) among $15 \times 15$ pixels centered at HEALPix nodes (i.e., locations of injected stars).

Internal diagnostics only reflect how {\sc Imcom} thinks it does; to gauge how it actually does, we use simulated layers with well-known properties. The second set of two criteria are based on simulated noise frames. In \paptwo\ and \papthree, the 2D noise power spectra are computed as
\begin{equation}
    P_{\rm 2D}(u, v) = \frac{s_{\rm out}^2}{N^2} \left| \sum_{j_x, j_y} S_{j_x, j_y} e^{-2\pi i s_{\rm out} (uj_x+vj_y)} \right|^2,
    \label{eq:P2D}
\end{equation}
where noise signal $S$ is the noise signal, $u$ and $v$ are sampled at integer multiples of $1/(Ns_{\rm out})$; here $N = n_1 n_2 = 2560$ is the number of pixels on each side of a block, and $s_{\rm out} \equiv \Delta\theta = 0.0390625 \,{\rm arcsec}$ is the output pixel scale.\footnote{After Equation~(19) of \papthree, it should read ``$s_{\rm out} \equiv \Delta\theta = 0.025 \,{\rm arcsec}$.'' We apologize for this typo.} Then the azimuthally averaged power spectra $P_{\rm 1D}(\upsilon)$ are computed using the method from \citet{2023MNRAS.520.4715C}: We take the azimuthal average of 2D power spectra over $150$ thin annuli of equal width. To facilitate comparisons, we integrate each 1D noise power spectrum to obtain the total noise power
\begin{equation}
    P_{\rm tot} = \int_0^{\upsilon_{\rm max}} 2\pi \upsilon P_{\rm 1D}(\upsilon) {\rm d}\upsilon,
    \label{eq:P1D_int}
\end{equation}
where $\upsilon_{\rm max} = 1/(\sqrt{2} s_{\rm out})$ is the maximum amplitude that can be measured from a coadded noise frame. The two noise-based evaluation criteria are simply
\begin{enumerate}
    \setcounter{enumi}{4}
    \item Total white noise power: $P_{\rm tot} [{\tt whitenoise10}]$;
    \item Total $1/f$ noise power: $P_{\rm tot} [{\tt 1fnoise9}]$.
\end{enumerate}
For each kind of noise, there is only one total power associated with each block; however, since a block has $2560^2$ pixels (not including padding), these measurements are not limited by statistical uncertainties.

The third set of five criteria are based on injected stars ({\tt \textquotesingle gsstar14\textquotesingle}). Our ultimate goal is to measure galaxies; compared to extended sources, point sources are narrower in real space and thus wider in Fourier space, hence we consider measurement of simulated stars as a ``stress test.'' Specifically, we use the {\sc HSM} module \citep{2003MNRAS.343..459H, 2005MNRAS.361.1287M} of {\sc GalSim}. To perform such measurements, a $3.086 \times 3.086 \,{\rm arcsec}^2$ ($79 \times 79$ output pixels) cutout is made for each of the injected point sources in each band. Specifically, the five evaluation criteria are:
\begin{enumerate}
    \setcounter{enumi}{6}
    \item Amplitude $A$, which corresponds to the zeroth moment;
    \item Centroid offset $d = \sqrt{d_x^2 + d_y^2}$, which corresponds to the first moments;
    \item Shear invariant width $s$, defined as
    \begin{equation}
    \label{eq:size_definition}
        s = \sqrt[4]{M_{xx}M_{yy}-M_{xy}^2};
    \end{equation}
    \item Ellipticity $g = \sqrt{g_1^2 + g_2^2}$, where the components are defined as
    \begin{equation}
        \label{eq:shape_definition}
        (g_1,g_2) = \frac{(M_{xx}-M_{yy}, 2M_{xy})}
        {M_{xx} + M_{yy} + 2\sqrt{M_{xx}M_{yy} - M_{xy}^2}};
    \end{equation}
    \item The spin-2 fourth moment $|M^{\rm (4)}_{\rm PSF}| = (\Re[M^{\rm (4)}_{\rm PSF}]^2 + \Im[M^{\rm (4)}_{\rm PSF}]^2)^{1/2}$,\footnote{In \papthree, this expression and its variants were missing the square root. We apologize for this imprudence.} where the complex version defined as
    \begin{equation}
        \label{eq:spin-2-fourth-moment}
        M^{\rm (4)}_{\rm PSF} = M_{40} - M_{04} + 2i (M_{31}+M_{13}),
    \end{equation}
    with standardized higher moments defined following \citet{2023MNRAS.520.2328Z}
    \begin{equation}
        \label{eq:moment_define}
        M_{pq} = \frac{\int {\rm d}x \, {\rm d}y \, u^p \, v^q \, \omega(x,y)
        \, I(x,y)}{\int {\rm d}x \, {\rm d}y \, \omega(x,y) \, I(x,y) },
    \end{equation}
    where $(u,v)$ are transformed coordinates in which Equation~(\ref{eq:shape_definition}) vanishes and Equation~(\ref{eq:size_definition}) evaluates to $1$; $p$ and $q$ are integer indices, $\omega(x,y)$ is the adaptive weight function, and $I(x,y)$ is the image. See \citet{2023MNRAS.525.2441Z} for the significance of this quantity.
\end{enumerate}
For a noiseless injected star with normalized flux and known location, $A$ and $s$ are determined by the target output PSF $\Gamma$, while $d$, $g$, and $|M^{\rm (4)}_{\rm PSF}|$ are all expected to be zero. We note that HSM is not designed for photometric measurement and does not perform aperture correction. Therefore, the amplitudes reported by HSM should not be considered as serious photometric solutions. Nonetheless, since HSM is closely related to current shape measurement tools, we still think HSM photometry (the seventh criterion defined above) is a useful diagnostic of image quality.

Finally, given the huge amount of data that the Roman HLWAS will yield, the computational cost of image processing is also something we need to take into account. The last evaluation criterion is thus
\begin{enumerate}
    \setcounter{enumi}{11}
    \item Time consumption per block.
\end{enumerate}
Note that the time consumption is subject to fluctuations of the powerfulness of specific computational facilities and can only be used as a rough reference.

\section{Configuration of Tests} \label{sec:config}

\begin{figure}
    \centering
    \includegraphics[width=\columnwidth]{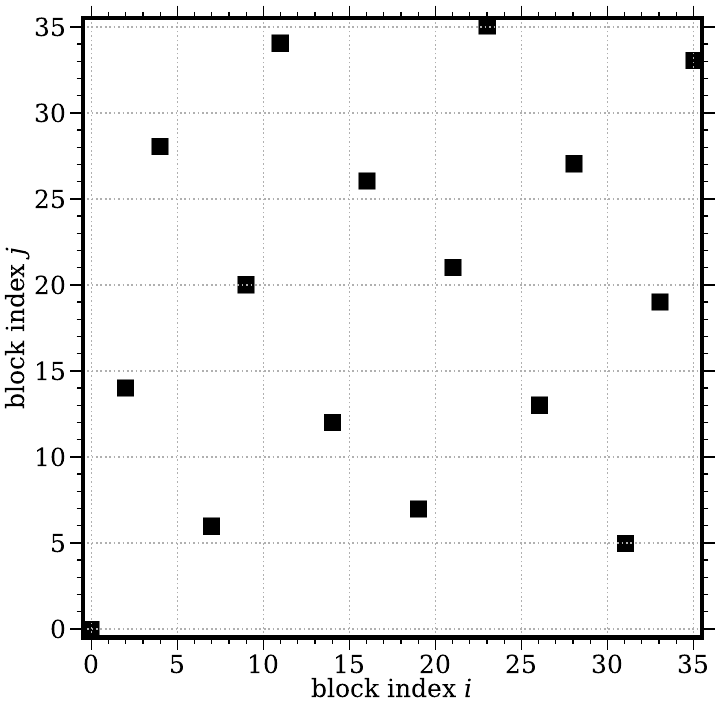}
    \caption{\label{fig:suite_diagram}Diagram showing the suite of $16$ test blocks ($1.75 \times 1.75 \,{\rm arcmin}^2$ each, including padding) selected from $36 \times 36$ blocks in the mosaic ($1.0 \times 1.0 \,{\rm deg}^2$). All cases studied in this work, benchmark or variant, are tested on these $16$ blocks.}
\end{figure}

\begin{table}[]
    \centering
    \caption{\label{tab:mean_coverage}Mean coverage information about the suite of $16$ selected blocks (see Figure~\ref{fig:suite_diagram}). Mean coverage is a rough estimate of how many input images overlap with the block. The second and third columns show the extrema among $16$ blocks, while the last column present the distribution in $5$ equally spaced bins.}
    \begin{tabular}{cccc}
    \hline
    Band & Minimum & Maximum & Histogram \\
    \hline
    Y106 & $3.9411$ & $5.9839$ & (2, 3, 2, 5, 4) \\
    J129 & $4.0838$ & $6.0000$ & (4, 5, 3, 2, 2) \\
    H158 & $4.0184$ & $5.9006$ & (4, 1, 5, 2, 4) \\
    F184 & $3.7633$ & $6.2591$ & (1, 3, 7, 3, 2) \\
    K213 & $4.2653$ & $6.0000$ & (3, 4, 2, 4, 3) \\
    \hline
    \end{tabular}
\end{table}

In this section, we detail the configuration of tests conducted in this work. As mentioned in Section~\ref{ss:ou24}, the {\sc Imcom} coadds included in OU24 data products were a $1.0 \times 1.0\,{\rm deg}^2$ mosaic ($36 \times 36$ blocks) in five bands. Since we want to test many different settings in this work, we choose to re-coadd a set of $16$ blocks with each configuration. Specifically, we choose blocks with indices $(i_{\rm block}, j_{\rm block}) = (691i // 36, 691i \% 36)$, where $//$ is integer division, $\%$ is the modulo operation, and $i = 0, 1...15$. Figure~\ref{fig:suite_diagram} illustrates the spatial distribution of these $16$ representative blocks. Furthermore, their mean coverages (i.e., numbers of contributing exposures) in each band are summarized in Table~\ref{tab:mean_coverage}; following \paptwo\ and \papthree, the binning shown here will be used for noise diagnostics.

\begin{table*}[]
    \centering
    \caption{\label{tab:var_config}Summary of variant settings studied in this work. Each variant case differs from the benchmark case of the corresponding linear algebra kernel in one and only one aspect. The third column specifies which kernel each variant applies to; ``both'' means both Cholesky and iterative kernels. All variants apply to all the five bands studied in this work (Y106, J129, H158, F184, and K213). See Section~\ref{sec:config} for further explanations and specific values.}
    \begin{tabular}{cccl}
    \hline
    Reference & Variant(s) & Kernel(s) & Description \\
    \hline
    \multirow{2}{*}{Section~\ref{ss:outpsf-m}} & {\tt airyobsc}, {\tt airyunobsc} & Both & Using Airy disks instead of Gaussians as target output PSFs \\
    & {\tt gauss\_0.8x}, {\tt gauss\_1.2x} & Both & Using Gaussians of different widths at target output PSFs \\
    \hline
    \multirow{4}{*}{Section~\ref{ss:kernel-r}} & {\tt kappac\_3x}, {\tt kappac\_9x} & Cholesky & Using larger Lagrange multipliers in the objective function \\
    & {\tt inpad=1.00}, {\tt inpad=0.76} & Cholesky & Using smaller acceptance radii for selecting input pixels \\
    & {\tt rtol=4.5e-3}, {\tt rtol=5.0e-4} & Iterative & Using different relative tolerances for linear system solving \\
    & {\tt inpad=0.75}, {\tt inpad=0.45} & Iterative & Using different acceptance radii for selecting input pixels \\
    \hline
    \multirow{4}{*}{Section~\ref{ss:exfeat-m}} & {\tt psfcirc} & Both & Applying circular cutouts to PSFs after sampling \\
    & {\tt psfnorm} & Both & Unifying normalization of PSFs after sampling \\
    & {\tt amppen} & Both & Penalizing large-amplitude Fourier modes in PSF leakage \\
    & {\tt flatpen} & Both & Penalizing flat-field differences between input images \\
    \hline
    \end{tabular}
\end{table*}

In each band, the benchmark case of the Cholesky kernel is configured following OU24 coadds, and that of the iterative kernel is configured with necessary adjustments (see Table~\ref{tab:base_config}). For each combination of bands and kernels, we explore $12$ variants, which are summarized in Table~\ref{tab:var_config}. We discuss choice of target output PSFs, kernel-specific settings, and some experimental features in Sections~\ref{ss:outpsf-m}, \ref{ss:kernel-m}, and \ref{ss:exfeat-m}, respectively.

\subsection{Choice of Target Output PSFs} \label{ss:outpsf-m}

As explained in Section~\ref{ss:recap}, {\sc Imcom} allows users to specify target output PSFs, which the as-realized coadded PSFs will resemble to the largest extent. Choosing appropriate target PSFs, including their functional form and size, is crucial for the quality of {\sc Imcom} outputs.

Since the optical part of a Roman PSF (i.e., an ``input'' PSF for {\sc Imcom}) is mainly an obstructed Airy disk \citep[e.g.,][]{Rivolta1986ApOpt} with diffraction spikes, it is natural to consider the convolution of an Airy disk with a Gaussian of width $\sigma$ as the target output PSF:
\begin{equation}
    \Gamma_{\rm Airy}({\boldsymbol s}) = \int_{{\mathbb R}^2}
    \frac{[J_1(\pi s'/\xi) - \varepsilon J_1(\pi \varepsilon s'/\xi)]^2}{\pi (1-\varepsilon^2) s'{^2}}
    \frac{{\rm e}^{-({\boldsymbol s}-{\boldsymbol s}')^2/(2\sigma^2)}}{2\pi\sigma^2}
    {\rm d}^2{\boldsymbol s}',
    \label{eq:airy}
\end{equation}
where $\xi=\lambda/D$ is the diffraction scale defined as the central wavelength (different for each filter) divided by the diameter of the Roman entrance pupil, $J_1$ denotes the Bessel function of the first kind, and $\varepsilon = 0.31$ is the linear obstruction fraction of the telescope. This was adopted by \papone; the $\sigma$ parameter, referred to as ``extra smooth,'' was chosen empirically for each band. With such a functional form, the obstructed Airy disk is the common component of input and output PSFs, which simplifies the derivation of expected noise power spectra (see Appendix~A of \paptwo).

However, this component is not a necessary one, as {\sc Imcom} can aim for any target PSF that is sufficiently wide. (Otherwise, {\sc Imcom} would encounter the ``division-by-zero'' problem in Fourier space; see Section~5 of \papthree.) For example, we can set $\varepsilon$ in Equation~(\ref{eq:airy}) to zero to ``eliminate'' obstruction caused by the secondary mirror of Roman. More importantly, during the development of {\sc PyImcom} and the preparation for OU24 coadds, we found that using a simple Gaussian function
\begin{equation}
    \Gamma_{\rm Gauss}({\boldsymbol s}) = \frac{{\rm e}^{-{\boldsymbol s}^2/(2\sigma^2)}}{2\pi\sigma^2}
    \label{eq:gauss}
\end{equation}
as the target output PSF has some benefits. Intuitively, although a Gaussian has large-wavenumber modes that do not exist in input images, our analysis pipelines usually work better with Gaussian or Gaussian-like PSFs. Therefore, both \papthree\ and OU24 used Gaussian PSFs.

\begin{figure*}
    \centering
    \includegraphics[width=\textwidth]{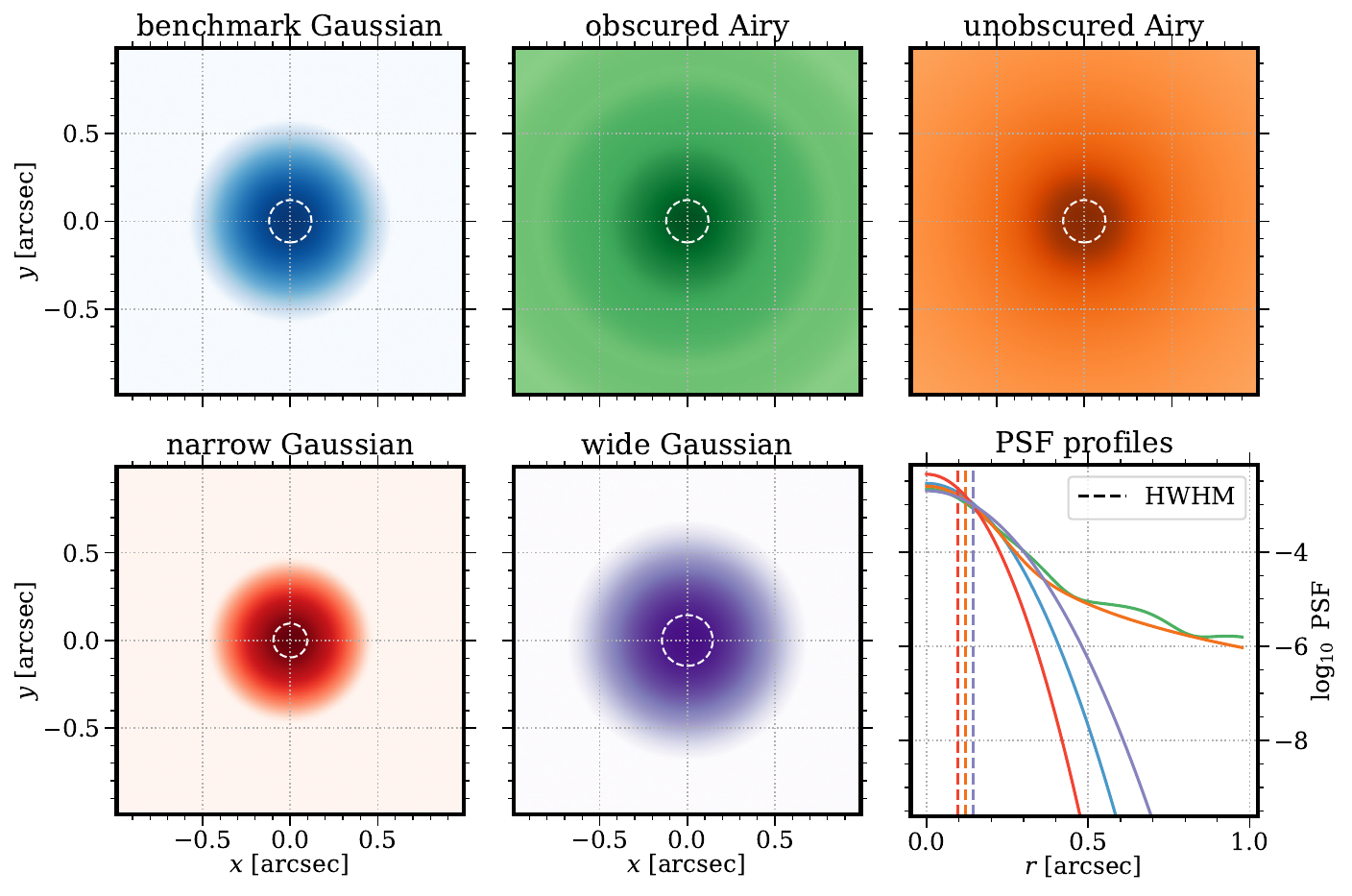}
    \caption{\label{fig:target_PSFs}Target output PSFs in the H158 band studied in this work. The normalization of PSFs is arbitrary but unified. The upper row shows three PSFs of the same FWHM but different forms; from left to right: Gaussian PSF in the benchmark case, obscured Airy disk convolved with a Gaussian, and unobscured Airy disk convolved with a Gaussian. Parameters of these three PSFs are tabulated in Table~\ref{tab:extrasmooth}. The left two panels of the lower row show Gaussian PSFs that are narrower and wider (in terms of FWHM) than the benchmark version by $20\%$, respectively. The lower right panel compares the radial profiles of these PSFs, shown as solid curves with colors corresponding to preceding panels. The half widths at half maximum (HWHMs) are shown as white dashed circles in the first five panels and colored dashed vertical lines in the last one.}
\end{figure*}

\begin{table}[]
    \centering
    \caption{\label{tab:extrasmooth}Full widths at half maximum (FWHMs) of target output PSFs studied in this work. All values are quoted in units of native pixel size (${\tt s\_in} = 0.11 \,{\rm arcsec}$). Values for narrower and wider Gaussian PSFs are not included as they can be easily calculated from those of benchmark Gaussian PSFs, of which the FWHMs are shown in the second column of this table. For each Airy PSF, either obscured or unobscured, two values are quoted: $\theta_{\rm Airy}$, FWHM of the Airy disk based on the geometry of the Roman Space Telescope and the central wavelength in that band; $\theta_{\rm Gauss}$, FWHM of the Gaussian component being convolved with the Airy disk (see Equation~(\ref{eq:airy})), so that the FWHM of the resulting PSF matches that of the benchmark Gaussian PSF in that band. See Figure~\ref{fig:target_PSFs} for an illustration of target output PSFs in the H158 band.}
    \begin{tabular}{c|c|cc|cc}
    \hline
    \multirow{2}{*}{Band} & Gaussian & \multicolumn{2}{c}{obscured Airy} & \multicolumn{2}{|c}{unobscured Airy} \\
    & $\theta_{\rm Gauss}$ & $\theta_{\rm Airy}$ & $\theta_{\rm Gauss}$ & $\theta_{\rm Airy}$ & $\theta_{\rm Gauss}$ \\
    \hline
    Y106 & $2.0$ & $0.8158$ & $1.7103$ & $0.8582$ & $1.7999$ \\
    J129 & $2.1$ & $0.9987$ & $1.7559$ & $1.0506$ & $1.8339$ \\
    H158 & $2.2$ & $1.2227$ & $1.7925$ & $1.2862$ & $1.8355$ \\
    F184 & $2.3$ & $1.4242$ & $1.8245$ & $1.4982$ & $1.8290$ \\
    K213 & $2.4$ & $1.6482$ & $1.8255$ & $1.7339$ & $1.7801$ \\
    \hline
    \end{tabular}
\end{table}

In this paper, we conduct tests to investigate the impact of both functional form and size of a target output PSFs. Specifically, we compare: i) Gaussian, obscured Airy, and unobscured Airy with the same width, and ii) Gaussian PSFs with three different widths. Figure~\ref{fig:target_PSFs} illustrate the five target PSFs in the H158 band; width parameters for the first set of comparisons are tabulated in Table~\ref{tab:extrasmooth}. (For the second set, the widths of narrow and wide Gaussians can be easily computed from those of benchmark Gaussians.) We note that for our benchmark cases, the widths of Gaussian PSFs are taken from OU24 coadds. As we can see from Figure~\ref{fig:target_PSFs}, compared to Airy disks, Gaussian PSFs are much more concentrated in real space, and thus extended in Fourier space. Linear obscuration in Equation~(\ref{eq:airy}) manifests as an oscillating feature in the radial profile; but because of the Gaussian smoothing, the obscured Airy disk still has a monotonic profile. From Table~\ref{tab:extrasmooth}, it is clear that $\theta_{\rm Airy}$ increases with the central wavelength of a band, and obscuration makes it smaller. Since the widths of the Gaussian PSFs were set empirically, it is not unexpected that the ``extra smooth'' factors ($\theta_{\rm Gauss}$) for Airy disks are sometimes not monotonic.

\subsection{Kernel-specific Hyperparameters} \label{ss:kernel-m}

To make coadded PSFs defined in Equation~(\ref{eq:outpsf}) as close to target PSFs as possible, {\sc Imcom} builds and solves linear systems, and specific linear algebra strategies are referred to as ``kernels.'' While the elements of the systems matrices ${\mathbf A}$ and ${\mathbf B}$ in Equation~(\ref{eq:T-AB}) are fully determined by PSFs, other ingredients of the solving strategy are still subject to change. This work explores three such ingredients, $\kappa_\alpha$, ${\tt rtol}$, and ${\tt INPAD}$, which we discuss below.

\paragraph{Lagrange multiplier $\kappa_\alpha$} In Equation~(\ref{eq:T-AB}), the Lagrange multiplier $\kappa_\alpha$ sets the balance between reducing PSF leakage $U_\alpha$ and reducing noise amplification $\Sigma_\alpha$. In their Appendix, \citet{2011ApJ...741...46R} showed that both $U_\alpha$ and $\Sigma_\alpha$ are monotonic functions of $\kappa_\alpha$: A larger $\kappa_\alpha$ leads to a smaller $\Sigma_\alpha$ at the expense of a larger $U_\alpha$, and vice versa. For OU24 coadds, which were produced by the Cholesky kernel, $\kappa_\alpha/C$ (where $C = \Vert\Gamma\Vert^2$, $\Gamma$ being the target output PSF) was uniformly set to $2 \times 10^{-4}$, $4 \times 10^{-4}$, $6 \times 10^{-4}$, $8 \times 10^{-4}$, and $1 \times 10^{-3}$ in the Y106, J129, H158, F184, and K213 bands, respectively. As can be seen from Equation~(\ref{eq:T-AB}), $\kappa_\alpha$ stabilizes linear systems by directly enlarging the diagonal terms of the matrices to be inverted; smaller $\kappa_\alpha$ values would destabilize the systems, so we only try larger ones in this work. Specifically, for the Cholesky kernel, we try two values in each band, which are larger than the benchmark value by factors of $3$ and $9$, respectively. As for the iterative kernel, we always set $\kappa_\alpha$ to zero, which was shown to be a reasonable choice in \papthree.

\paragraph{Relative tolerance ${\tt rtol}$} As an iterative algorithm, the conjugate gradient method needs a tolerance parameter to set the stop condition. For a general linear system $A\vec{x} = \vec{b}$, the relative error is $\varepsilon = |\vec{b} - A\vec{x}| / |\vec{b}|$, where $|\cdot|$ denotes the Euclidean norm; the relative tolerance ${\tt rtol}$ is the maximum allowed value of $\varepsilon$. In \papthree, we chose ${\tt rtol} = 1.5 \times 10^{-3}$, which unfortunately lead to significant random errors in coaddition weights, dubbed ``output white noise.'' In this paper, we use $1.5 \times 10^{-3}$ for benchmark cases, and try a larger value and a smaller one, both by a factor of $3$, to gauge how the quality of the outputs scales with this parameter.

\paragraph{Acceptance radius ${\tt INPAD}$} For each postage stamp (in the case of the Cholesky kernel) or each output pixel (in the case of the iterative kernel), we need to select a set of input pixels in its vicinity; the acceptance radius ${\tt INPAD}$ specifies what ``vicinity'' exactly means. The number of input pixels $n$ is a quadratic function of ${\tt INPAD}$ for both kernels (see Equations~(8) and (17) in \papthree, where $\rho_{\rm acc} \equiv {\tt INPAD}$); meanwhile, the complexity of linear system solving is either ${\cal O}(n^3/6 + n^2m)$ (Cholesky) or ${\cal O}(n^2m)$ (iterative), hence ${\tt INPAD}$ is expected to significantly affect the time consumption. We note that the number of input pixels is proportional to the number of contributing input images, hence the time consumption scales differently with coverage for the two linear algebra kernels. However, the HLIS coverage is largely uniform ($6$ images minus detector gaps, bad pixels, cosmic rays hits, etc.), therefore ${\tt INPAD}$ is the decisive factor for computational complexity. In \papthree, specifically the lower panel of Figure~3 and the entire Figure~4, we demonstrated that although the ${\mathbf T}$ matrices are not strictly sparse, elements corresponding to input pixels far away from a given output pixel $\alpha$ are usually small. Combining these two factors, we have both motivation and justification for trying smaller ${\tt INPAD}$ values. For the Cholesky kernel, we try $1.00$ and $0.76$ in addition to the benchmark $1.24 \,{\rm arcsec} \approx 11.27 \,{\tt s\_in}$, ${\tt s\_in} = 0.11 \,{\rm arcsec}$ being the input pixel scale. As for the iterative kernel, it is a legitimate concern that the benchmark $0.60 \,{\rm arcsec} \approx 5.45 \,{\tt s\_in}$ is too small, hence we try $0.75$ and $0.45$ as additional values.

\subsection{Design of Experimental Features} \label{ss:exfeat-m}

To address other practical issues involved in the coaddition process, we have come up with some experimental features that may lead to improvements. In this paper, we investigate four of them, which we now describe.

\paragraph{Circular cutouts for PSF arrays ({\tt psfcirc})} Mathematically, a point spread function (PSF) is a two-dimensional probability density function describing the landing location of a detected photon; it is defined in ${\mathbb R}^2$, although it vanishes at large distances. Yet in practice, we can only sample its values on a regular grid of pixels (see Section~B.2 of \papthree\ for some details about how {\sc PyImcom} handles this). Such sampling introduces two preferred directions in ${\mathbb R}^2$, namely the $x$ and $y$ directions, which are not a part of the nature of the PSF. Therefore, it is potentially beneficial to make a circular cutout for each PSF array. In {\sc PyImcom}, this is implemented as only allowing non-zero values in an approximate circular region surrounding the center, of which the radius is half the side length of the square array.

\paragraph{Unified normalization of PSF arrays ({\tt psfnorm})} In addition to being square, the PSF sampling arrays are also finite. Consequently, if we integrate PSFs over the sampled (square) region, the results are usually smaller than $1$, and the discrepancies from $1$ vary from PSF to PSF. Note that such discrepancies are larger for Airy disks than Gaussian functions (see Figure~\ref{fig:target_PSFs}). To try to treat input images equally, a plausible way is to normalize all as-sampled PSF arrays to a same value. The specific normalization does not matter, as ${\mathbf A}$ and ${\mathbf B}$ matrices in Equation~(\ref{eq:T-AB}) have the same units.

\paragraph{Penalizing large-amplitude modes ({\tt amppen})} While computing overlaps between PSF arrays and thus elements of system matrices, it is possible to assign different weights to different modes in Fourier space. The weighting function, $\tilde\Upsilon({\boldsymbol u})$ (see Equations~(17) to (19) of \citet{2011ApJ...741...46R}), was originally chosen to be uniform ($\tilde\Upsilon({\boldsymbol u}) = 1$), yet \papone\ accidentally used $\tilde\Upsilon({\boldsymbol u}) = [1 + A_\Upsilon {\rm e}^{-2\pi^2\sigma_\Upsilon^2(u^2+v^2)}]^2$ with $A_\Upsilon=1$ and $\sigma_\Upsilon = \frac32\times 0.11 \,{\rm arcsec}$. This places more weight on the small-amplitude (low-frequency) modes relative to the large-amplitude (high-frequency) modes, and is thus referred to as ``amplitude penalty.'' It is important to note that adopting a non-uniform weighting is different from choosing another target output PSF, as the latter only affects the ${\mathbf B}$ matrix in Equation~(\ref{eq:T-AB}), while the former also affects ${\mathbf A}$. We try the same values as in \papone\ to see if such penalty is worth including.

\paragraph{Penalizing flat-field differences ({\tt flatpen})} To attenuate negative effects of differences in flat-field levels, \papone\ also penalized unequal contributions (i.e., total weights in the ${\mathbf T}$ matrix) from different exposures. This is done by adding a correction term
\begin{equation}
    \Delta A_{ij} = {\tt flat\_penalty} \times (\delta_{\bar{i}\bar{j}} - 1/\bar{n})
    \label{eq:flatpen}
\end{equation}
to the ${\mathbf A}$ matrix, where the {\tt flat\_penalty} parameter is chosen to be $2 \times 10^{-7}$ in this work, $\delta$ is the Kronecker delta, $\bar{i}$ ($\bar{j}$) is the index of parenting image of input pixel $i$ ($j$), and $\bar{n}$ denotes the number of relevant input images. Likewise, we conduct tests to see if this penalty is worth including.

Before we conclude this section, we note that all settings mentioned above are configurable via the {\sc PyImcom} interface, and configuration files for all tests in this paper are available on GitHub.\footnote{\url{https://github.com/Roman-HLIS-Cosmology-PIT/pyimcom/tree/main/configs/paper4_configs}}

\section{Benchmark Results} \label{sec:base}

Before investigating how {\sc Imcom} hyperparameters affect its outputs in the next section, we first establish our benchmark results in this section.

\begin{figure*}
    \centering
    \includegraphics[width=\textwidth]{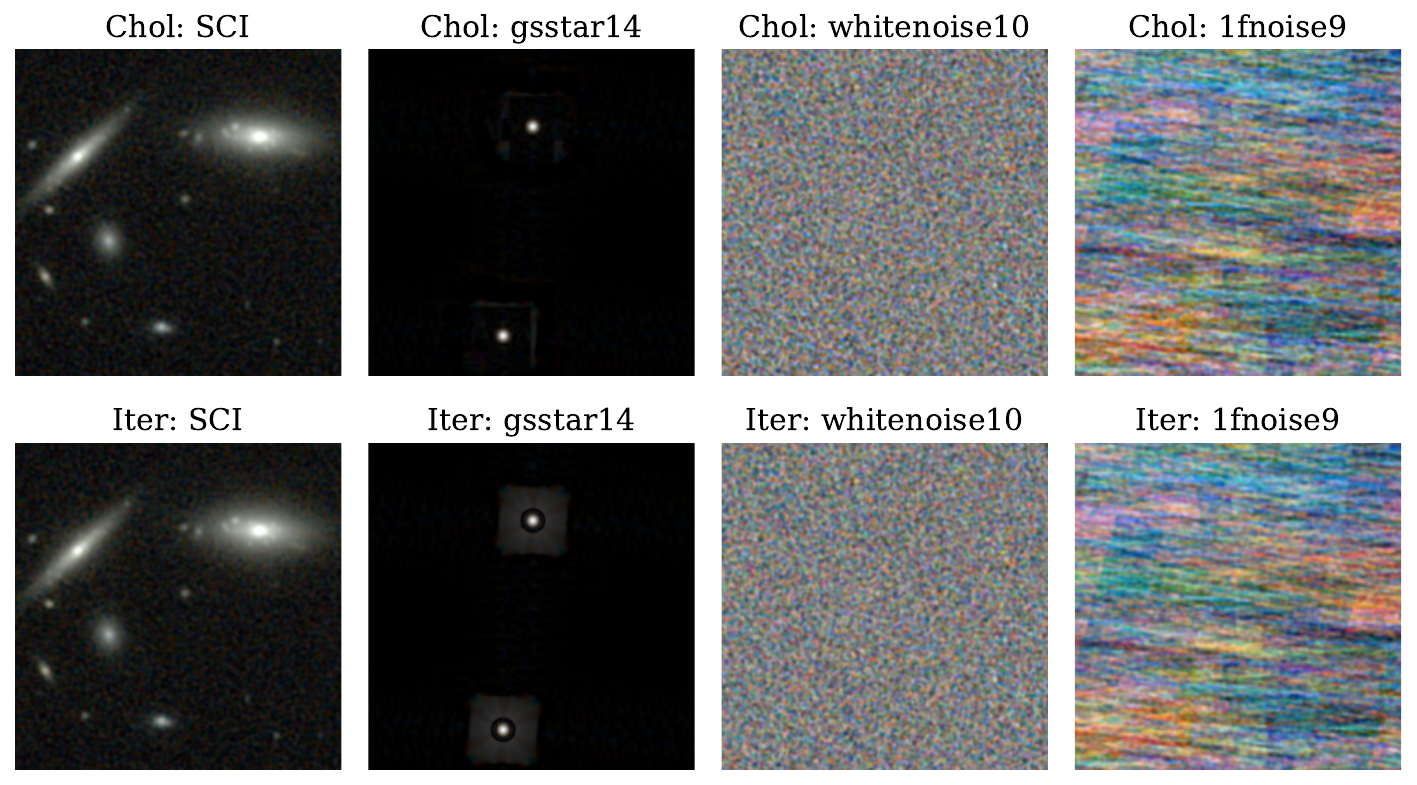}
    \caption{\label{fig:example_images}Four layers in a field of $17.5 \,{\rm arcsec}$ ($448$ output pixels) on a side, coadded by the Cholesky kernel (upper row) and the iterative kernel (lower row). Each panel is a Y106 ({\color{Y106} \#001AA6}) + J129 ({\color{J129} \#006659}) + H158 ({\color{H158} \#596600}) + F184 ({\color{F184} \#A61A00}) composite; note that these four colors have similar lightnesses and add up to white (\#FFFFFF). From left column to right column, the four layers are: simulated science images ({\tt \textquotesingle SCI\textquotesingle}), injected stars drawn by {\sc GalSim} ({\tt \textquotesingle gsstar14\textquotesingle}), simulated white noise frames ({\tt \textquotesingle whitenoise10\textquotesingle}), and simulated $1/f$ noise frames ({\tt \textquotesingle 1fnoise9\textquotesingle}). The scaling is set following \papone\ Figure~8 for {\tt \textquotesingle SCI\textquotesingle} and following \paptwo\ Figure~1 for the other three layers.}
\end{figure*}

Similar to Figure~5 in \papthree, Figure~\ref{fig:example_images} compares four layers of a $17.5 \times 17.5 \,{\rm arcsec}^2$ field produced by the benchmark cases of the Cholesky and iterative kernels. Because of the difference in output pixel scale (see Table~\ref{tab:base_config}), the same area of the sky corresponds to less pixels in this work. Through visual inspection, it is hard to notice the differences between the two versions of simulated science images (first column) or noise frames (last two columns); we conclude that both linear algebra kernels can successfully coadd OU24 images, as expected. The {\sc GalSim} injected stars (second column) manifest different artifacts associated with the two kernels: postage stamp boundary effects (features in the $x$ and $y$ directions) for the Cholesky kernel and lingering input PSF stamps (corresponding to the roll angles of input images) for the iterative kernel. See Section~6 of \papthree\ for a fuller discussion.

\begin{figure*}
    \centering
    \includegraphics[width=\textwidth]{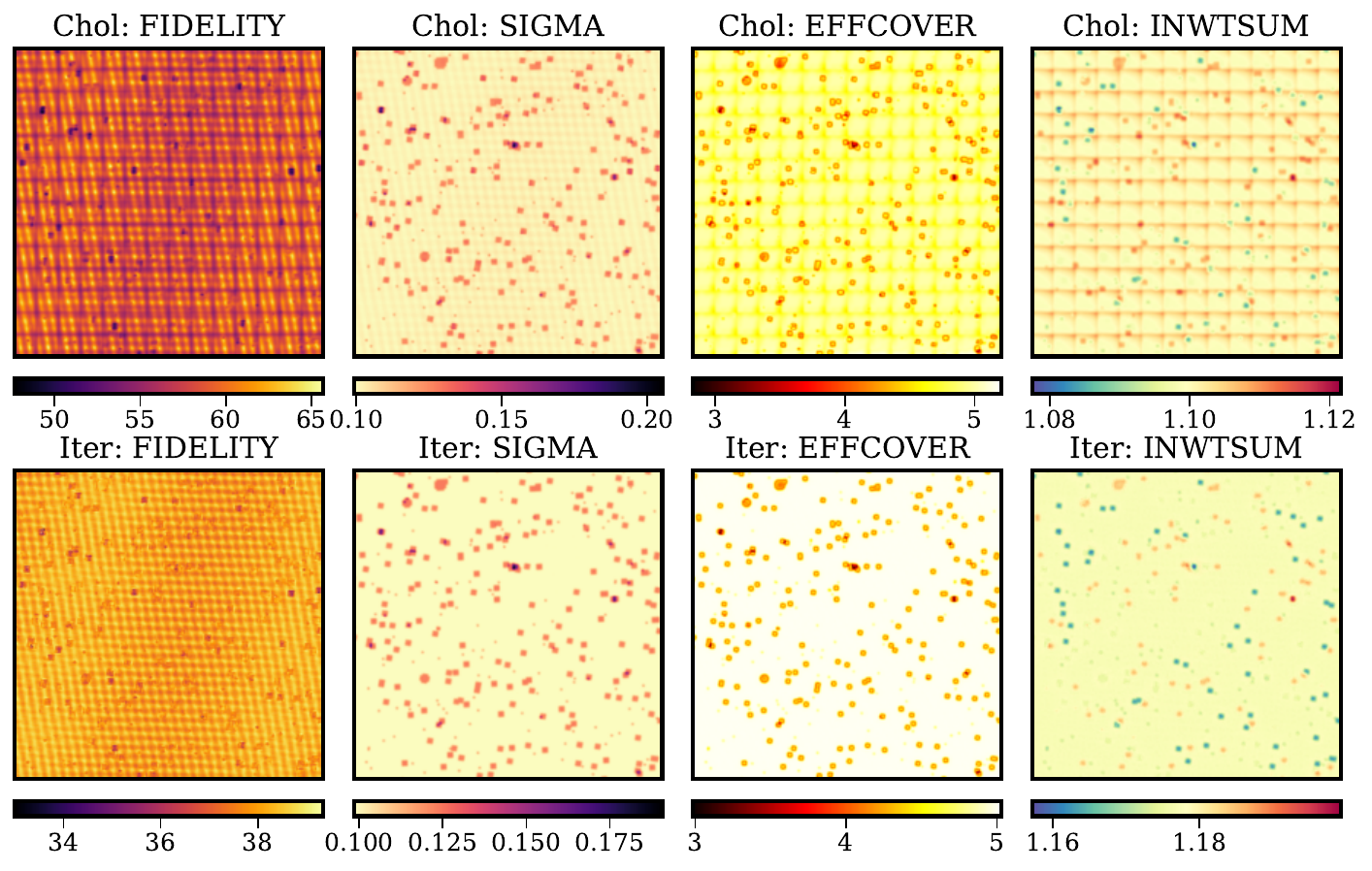}
    \caption{\label{fig:example_outmaps}Output maps in the H158 band produced by the Cholesky kernel (upper row) and the iterative kernel (lower row). From left to right: fidelity (negative logarithmic PSF leakage in decibels, i.e., $-10 \log_{10} (U_\alpha/C)$), noise amplification (Equation~(\ref{eq:U_Sigma})), effective coverage (Equation~(\ref{eq:Neff})), and total weight (Equation~(\ref{eq:Tsum})). Note that we deliberately choose different color bar ranges to better display spatial structures.}
\end{figure*}

Like Figure~6 in \papthree, Figure~\ref{fig:example_outmaps} displays {\sc Imcom} output maps in the same field as Figure~\ref{fig:example_images} in the H158 band. In the Cholesky kernel results, postage stamp boundaries can be seen in PSF fidelity and total input weight maps; they are much less obvious in the effective coverage map compared to in \papthree, likely due to a wider target output PSF (the FWHM was $1.9 \,{\tt s\_in}$ in \papthree\ and is $2.2 \,{\tt s\_in}$ in this work). They do not appear in the noise amplification map produced by the Cholesky kernel or any map produced by the iterative kernel, which agrees with what we saw in \papthree. Moir\'e patterns (see Section~5.3 of \papone\ for discussion) manifest as tilted grids in both fidelity maps and the Cholesky kernel noise amplification map. This small area of the sky does not contain any SCA boundary in the H158 band, which is likely because $17.5 \,{\rm arcsec} \approx 159 \,{\tt s\_in}$ and the SCA side length is $4088 \,{\tt s\_in} \approx 7.5 \,{\rm arcmin}$. Nevertheless, we can still see the impact of detector defects and cosmic ray hits on the ``actual'' coverage of each output pixel and its consequences.

\begin{figure*}
    \centering
    \includegraphics[width=\textwidth]{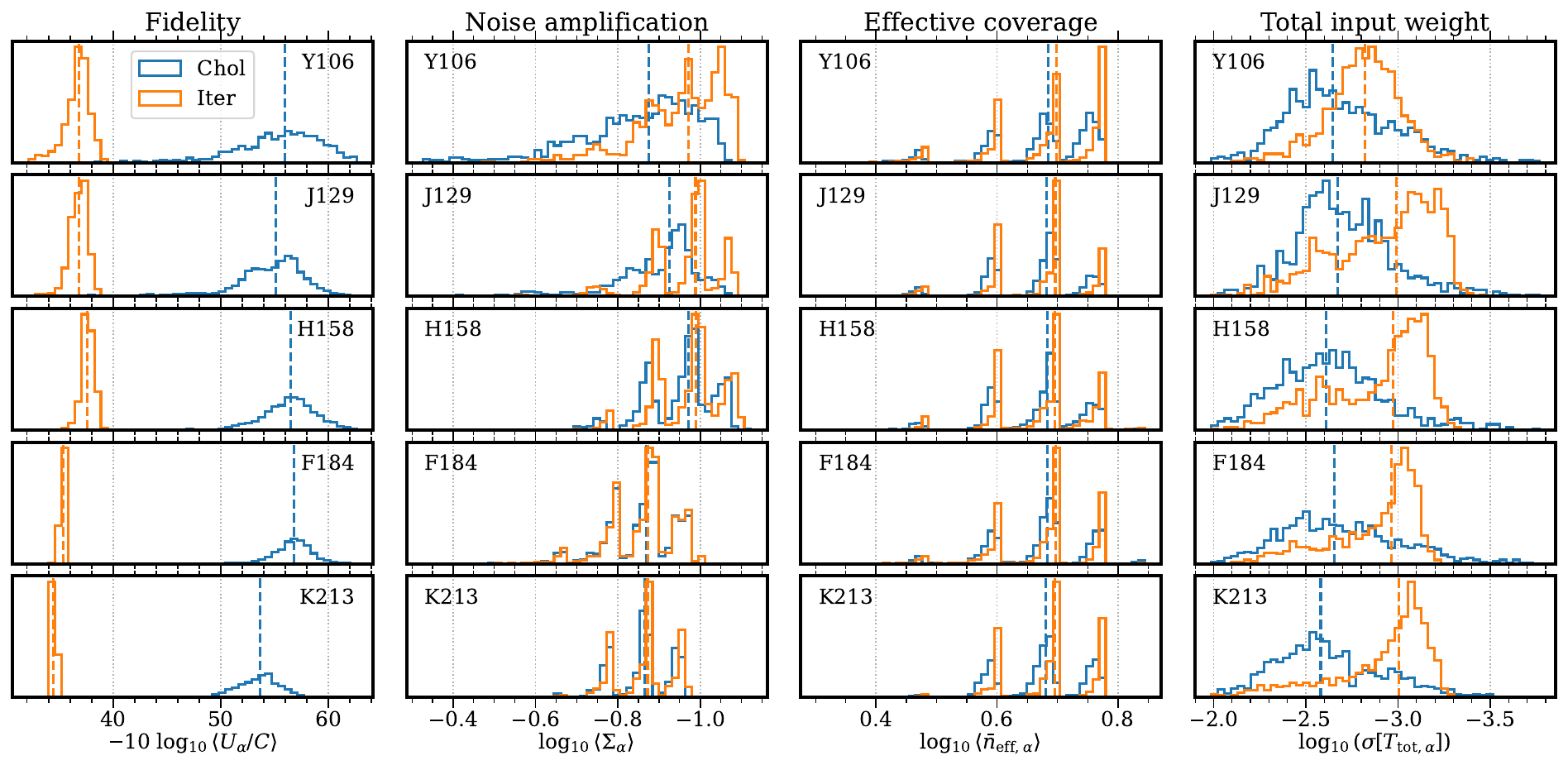}
    \caption{\label{fig:hist_outmap}Histograms of $957$ sets of four {\sc Imcom} diagnostics yielded by two linear algebra kernels in five bands. (From left to right:) Mean fidelity is defined as $-10\log_{10} \langle U_\alpha/C\rangle$, where $U_\alpha$ is the PSF leakage metric defined in Equation~(\ref{eq:U_Sigma}), and $\langle\cdot\rangle$ denotes an average over $15 \times 15$ pixels centered at a HEALPix node with ${\tt NSIDE} = 14$. Logarithmic mean noise amplification is defined as $\log_{10} \langle \Sigma_\alpha\rangle$, where $\Sigma_\alpha$ is the noise amplification metric defined in Equation~(\ref{eq:U_Sigma}). Logarithmic mean effective coverage is defined as $\log_{10} \langle \bar{n}_{{\rm eff}, \alpha}\rangle$, where $\bar{n}_{{\rm eff}, \alpha}$ is the effective coverage metric defined in Equation~(\ref{eq:Neff}). Logarithmic standard deviation of total weight is defined as $\log_{10} \sigma[T_{{\rm tot}, \alpha}]$, where $T_{{\rm tot}, \alpha}$ is the total input weight metric defined in Equation~(\ref{eq:Tsum}), and $\sigma[\cdot]$ denotes a standard deviation within $15 \times 15$ pixels centered at a HEALPix node with ${\tt NSIDE} = 14$. Following \papthree, we invert $x$-axes of the second and fourth columns so that ``better'' values are shown on the right; but unlike in \papthree, we do not explicitly introduce minus signs here. From top to bottom, the five rows present histograms in Y106, J129, H158, F184, and K213 bands; results given by the Cholesky and iterative kernels are shown in blue and orange, respectively.}
\end{figure*}

Figure~\ref{fig:hist_outmap} corresponds to the combination of Figures~7 to 10 in \papthree, with the removal of the empirical kernel and the addition of the K213 band. In terms of PSF fidelity (first column), with appropriately set target PSF sizes, the Cholesky kernel results in the F184 band are now consistent with those in the other bands. The fidelity values given by the Cholesky kernel are lower than those reported in \papthree, which can be ascribed to different coverages in these two samples (see legends of Figures~13 and 15 in \papthree\ and Table~\ref{tab:mean_coverage} in this paper). As for effective coverage (third column), we see clearly peaks corresponding a physical coverage of (from right to left) $6$, $5$, $4$, and $3$ and tails due to lost pixels in the iterative kernel results; the Cholesky kernel results are smeared towards low coverage, like in \papthree. The translation from coverage to noise amplification (second column) is at different levels in different bands, because of different PSF sizes: The larger the extent to which the target output PSF is wider than the input PSFs, the more dispersed the spatial distribution of coaddition weights (i.e., the more input pixels carrying ``significant'' weights), and thus the smaller the noise amplification. Unlike in \papthree, the iterative kernel is consistently better than the Cholesky kernel in terms of total input weight uniformity (last column). While the former is still limited by random errors, the latter is subject to postage stamp boundary effects (see the upper right panel of Figure~\ref{fig:example_outmaps}), which a cutout of $15 \times 15$ pixels is more likely to encounter when $n_2$ is $32$ (this work) rather than $50$ (\papthree).

\begin{figure*}
    \centering
    \includegraphics[width=\textwidth]{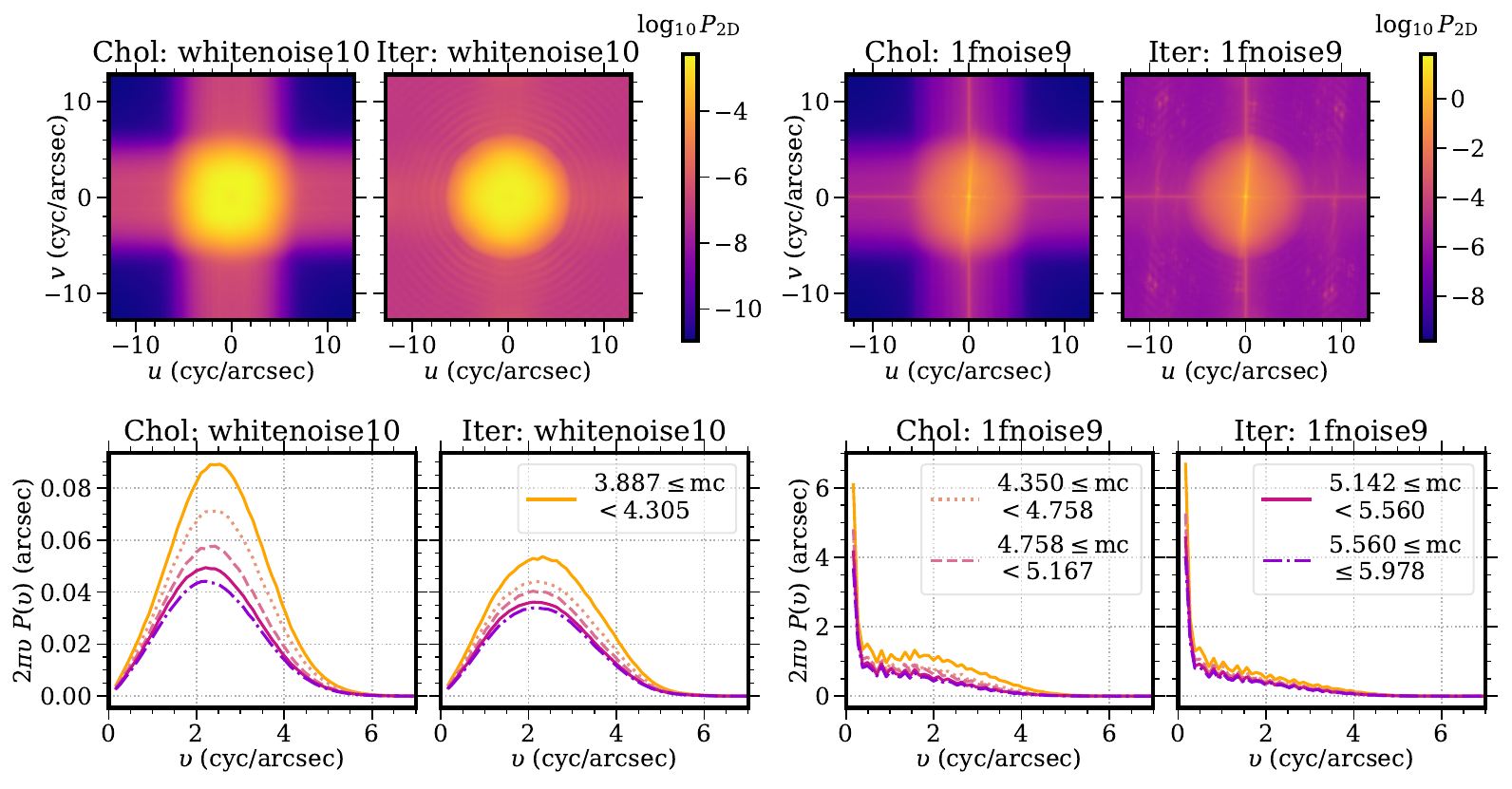}
    \caption{\label{fig:noiseps_Y106}Power spectra of simulated noise frames in the Y106 band. Upper row: Two-dimensional power spectra of simulated white noise frames ({\tt \textquotesingle whitenoise10\textquotesingle}) and $1/f$ noise frames ({\tt \textquotesingle 1fnoise9\textquotesingle}), averaged over $16$ test blocks (see Figure~\ref{fig:suite_diagram}) of each band-kernel combination and binned by $8 \times 8$ modes, plotted on a logarithmic scale. Following \paptwo\ Figure~2, the horizontal and vertical axes show wave vector components ($u$ and $v$ respectively) ranging from $-12.8$ to $+12.8$ cycles arcsec$^{-1}$; note that this range is set by the output pixel scale $\Delta\theta$ (see Table~\ref{tab:base_config}). The color scale shows the power $P(u, v)$ in units of ${\rm arcsec}^2$ (Equation~(\ref{eq:P2D})). Lower row: Azimuthally averaged power spectra of the same noise frames, averaged over modes within each of the $150$ radial bins and blocks in each mean coverage (``mc'' in short; see Table~\ref{tab:mean_coverage}) bin for each band-kernel combination. Results given by the Cholesky and iterative kernels are shown in odd and even columns, respectively.}
\end{figure*}

Figure~\ref{fig:noiseps_Y106} is a counterpart to Section~5 of \papthree, which was dedicated to the study of power spectra of coadded noise frames. With different simulated images and {\sc Imcom} settings, we recover all the features identified and discussed in \paptwo\ and \papthree: central bright regions mirroring the ``quotient'' of the target output PSF and the input PSFs, $+$ signs due to the choice of analysis method (Equation~(\ref{eq:P2D}); see \citet{2024PASP..136l4506L} for an alternative choice), ring features caused by the selection of input pixels ({\tt \textquotesingle whitenoise10\textquotesingle}), X shapes corresponding to the roll angles of input images, etc. We thus refer interested readers to our previous papers for fuller discussions. Figure~\ref{fig:noiseps_Y106} only shows results in the Y106 band; in the redder bands, the shapes of the two-dimensional power spectra are similar, and the Cholesky kernel gradually becomes almost as good as the iterative kernel in terms of noise control (also see the second column of Figure~\ref{fig:hist_outmap}). Here we emphasize that in the next section, the integral of noise power, Equation~(\ref{eq:P1D_int}), is computed for each curve in the second row of Figure~\ref{fig:noiseps_Y106} to show the impact of coverage.

\begin{figure*}
    \centering
    \includegraphics[width=\textwidth]{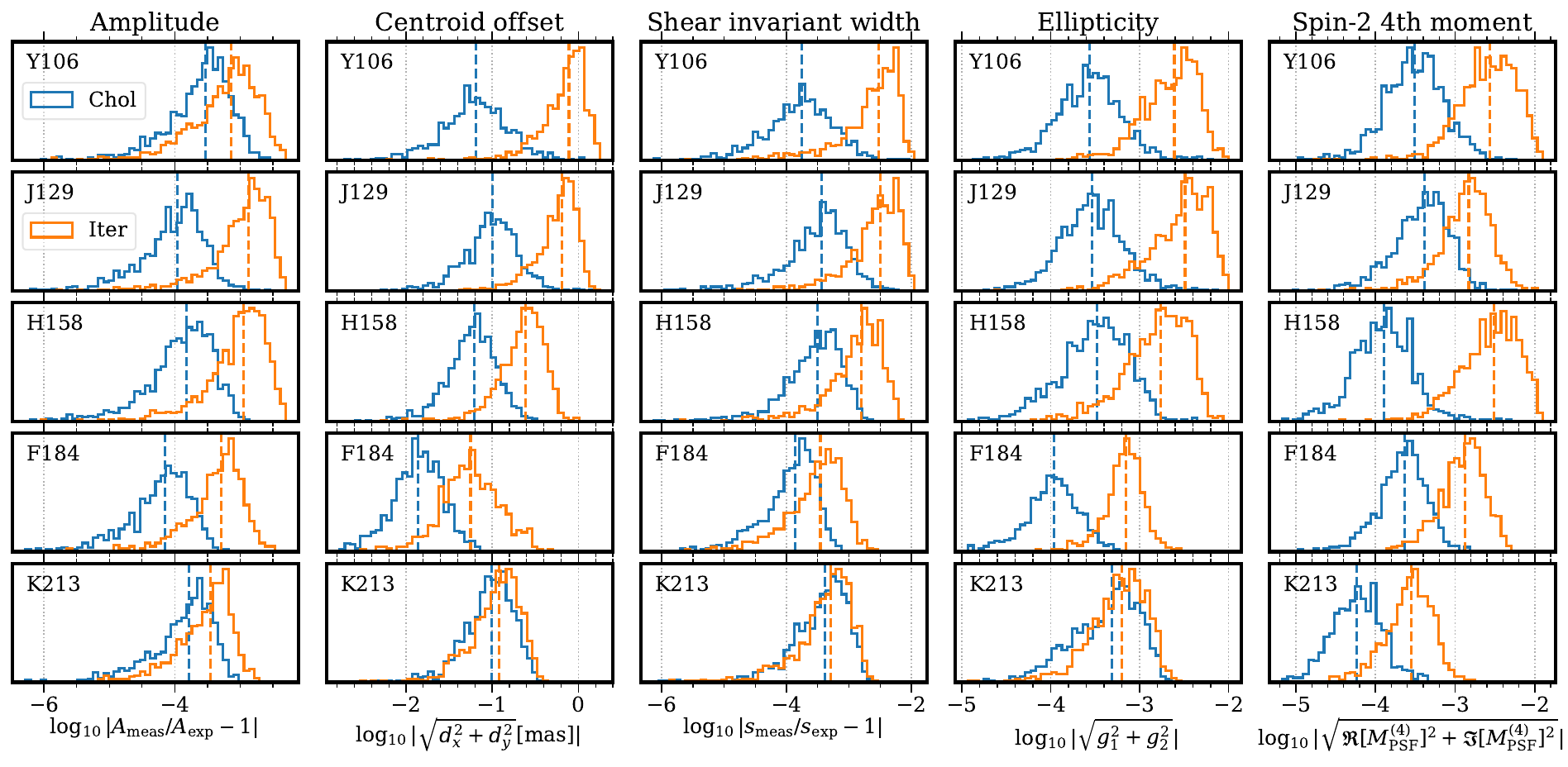}
    \caption{\label{fig:hist_star}Histograms of five measured properties of $957$ injected stars coadded by two linear algebra kernels in five bands. (From left to right:) Logarithmic absolute amplitude error is computed as $\log_{10} |A_{\rm meas}/A_{\rm exp}-1|$, where $A_{\rm meas}$ and $A_{\rm exp}$ are measured and expected amplitudes of an injected star, respectively. Logarithmic centroid offset is computed as $\log_{10} |\sqrt{d_x^2 + d_y^2}|$, where $d_x$ and $d_y$ are $x$ and $y$ components of the centroid offset (discrepancy between measured and expected centroids; in milliarcseconds) of an injected star, respectively. Logarithmic absolute size error is computed as $\log_{10} |s_{\rm meas}/s_{\rm exp}-1|$, where $s_{\rm meas}$ and $s_{\rm exp}$ are measured and expected shear-invariant widths of an injected star, respectively. Logarithmic ellipticity is computed as $\log_{10} \sqrt{g_1^2 + g_2^2}$, where $g_1$ and $g_2$ are the two components of measured ellipticity of an injected star. Logarithmic spin-$2$ fourth moment is computed as $\log_{10} (\sqrt{\Re[M^{\rm (4)}_{\rm PSF}]^2 + \Im[M^{\rm (4)}_{\rm PSF}]^2})$, where $\Re[M^{\rm (4)}_{\rm PSF}]$ and $\Im[M^{\rm (4)}_{\rm PSF}]$ are real and imaginary components of measured spin-$2$ fourth moment of an injected star, respectively. Note that both ellipticity and spin-$2$ fourth moment are expected to be zero for ideal, circular sources. The layout and format of these histograms are the same as those in Figure~\ref{fig:hist_outmap}. Unlike in Figure~\ref{fig:hist_outmap}, ``better'' values are shown on the left to be consistent with corresponding histograms in \papthree.}
\end{figure*}

Figure~\ref{fig:hist_star} corresponds to the combination of Figures~17 to 21 in \papthree. For all five moment-based measurements in all five bands, the Cholesky kernel outperforms the iterative kernel by about an order of magnitude, consistent with what we found in Section~6 of \papthree. The main reason is that the Cholesky kernel provides exact solutions to linear systems while the iterative kernel only provides approximate ones. For some of the measurements, the discrepancies are smaller in the K213 band, which will not be included in the ``Medium Tier'' and ``Wide Tier'' of Roman HLWAS \citep{2025arXiv250510574O}, hence it is a solid conclusion that the Cholesky kernel is the best known linear algebra strategy for {\sc Imcom}.

\begin{table}[]
    \centering
    \caption{\label{tab:time_consump}Number of cores and average time consumption (together with standard deviation) per block ($1.75 \times 1.75 \,{\rm arcmin}^2$) for the benchmark case of each linear algebra strategy in each band. Note that this work uses the same machine, namely the Pitzer cluster at the Ohio Supercomputer Center \citep{Pitzer2018}, as most of \papone\ simulations and all \papthree\ simulations.}
    \begin{tabular}{ccc}
    \hline
    Band & Cholesky & Iterative \\
    \hline
    No. of cores & $1.25$ & $1$ \\
    Y106 (hr) & $23.36 \pm 3.28$ & $33.62 \pm 4.08$ \\
    J129 (hr) & $18.32 \pm 5.96$ & $29.75 \pm 2.74$ \\
    H158 (hr) & $22.19 \pm 2.16$ & $28.26 \pm 3.01$ \\
    F184 (hr) & $24.93 \pm 3.64$ & $29.04 \pm 3.68$ \\
    K213 (hr) & $23.20 \pm 2.29$ & $30.78 \pm 2.88$ \\
    \hline
    \end{tabular}
\end{table}

Finally, we take a look at the time consumption of the benchmark cases as tabulated in Table~\ref{tab:time_consump}. We recall that the complexity of linear system solving with the Cholesky kernel is ${\cal O}(n^3/6 + n^2m)$, while that with the iterative kernel ${\cal O}(n^2m)$, a smaller $n_2$ (and thus a smaller $m$) should benefit the latter to a larger extent. Furthermore, since the Cholesky kernel uses a much larger acceptance radius (${\tt INPAD}$; see Table~\ref{tab:base_config}), it requires more memory, and we have to request $20$ cores for coadding $16$ blocks in parallel; the iterative kernel allows for single-core runs. However, even if we take this factor of $1.25$ into account, the Cholesky kernel is still less core-hour consuming than the iterative kernel in most bands, although the contrast is not as sharp as in Table~2 of \papthree. We reiterate that time consumption is subject to hardware fluctuations and is less reliable than other criteria.

\section{Variant Results} \label{sec:var}

With the benchmark results in mind, we now present the variant results in a much more condensed way. This section contains three figures of the same format: $12$ rows for the $12$ evaluation criteria defined in Section~\ref{ss:eval}, $5$ columns for the $5$ bands studied in this work; each panel contains either $10$ ``violins'' showing distributions of quantities (most criteria) or $10$ groups of data points showing values in different mean coverage bins (noise powers only). Note that we invert $y$-axes as needed so that desirable values are shown on the top of each row. Control over white noise (which is the main component of the readout noise) and ellipticity errors (which are directly related to shear errors) are arguably the two most important criteria, hence the corresponding numbers are tabulated in Appendix~\ref{app:tables}. The three subsections of this section correspond to those of Section~\ref{sec:config}, respectively.

\subsection{Target Output PSFs} \label{ss:outpsf-r}

\begin{figure*}
    \centering
    \includegraphics[width=0.9\textwidth]{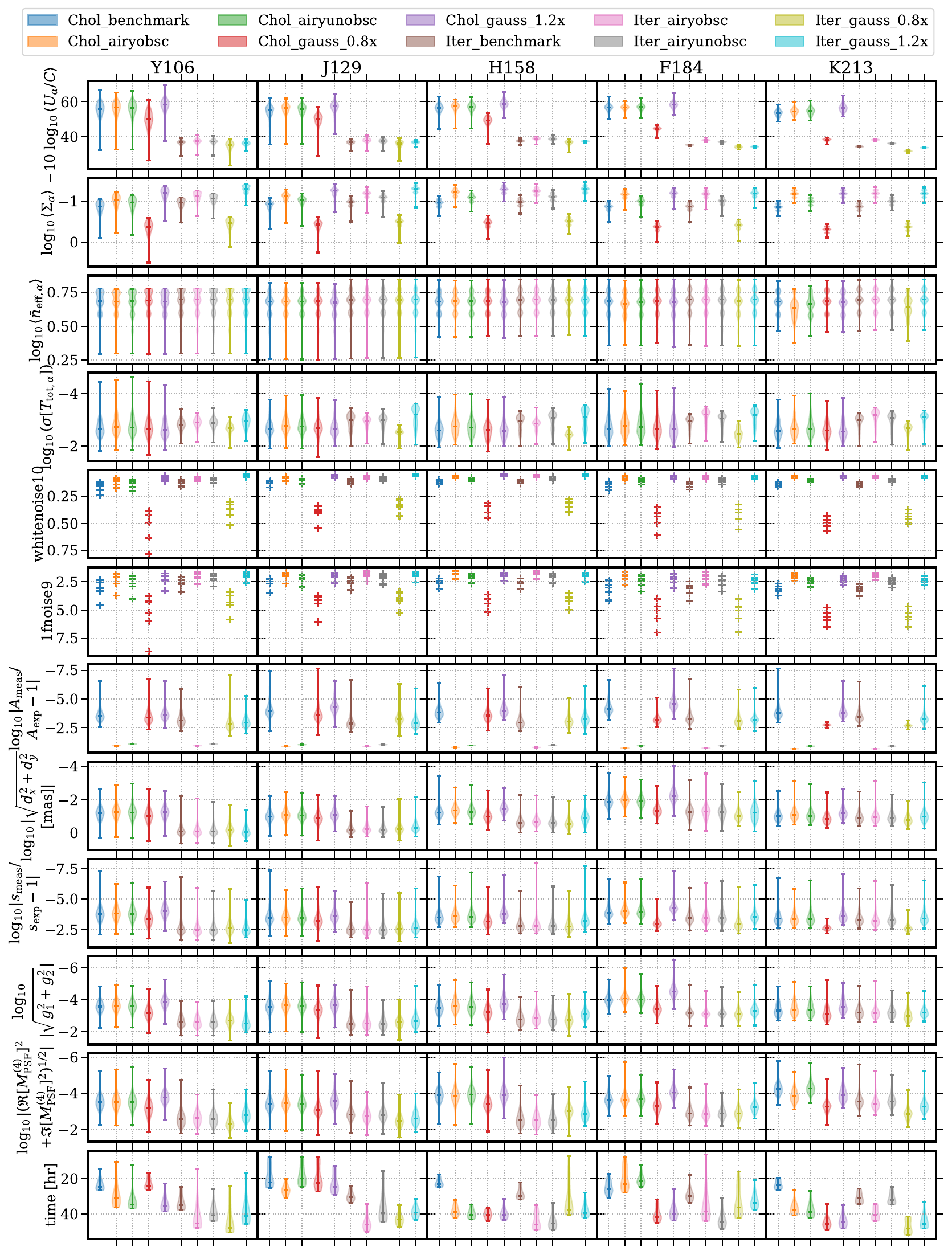}
    \caption{\label{fig:outpsf}Violin plots comparing different choices of target output PSFs. The five columns show results in the Y106, J129, H158, F184, and K213 bands, respectively. The first four rows present {\sc Imcom} diagnostics like those in Figure~\ref{fig:hist_outmap}; the fifth and sixth rows present integrals of azimuthally averaged power spectra like those the second row of Figure~\ref{fig:noiseps_Y106}; the seventh to eleventh rows present measured properties of injected stars like those in Figure~\ref{fig:hist_star}; the last row presents time consumption in hours similar to Table~\ref{tab:time_consump}. $y$-axes are inverted as needed so that desirable values are shown on the top of each row. Different colors correspond to different cases defined in Table~\ref{tab:var_config}.}
\end{figure*}

Figure~\ref{fig:outpsf} compares different target output PSFs introduced in Section~\ref{ss:outpsf-m}. We see two consistent trends across results in all five bands and with both the Cholesky and iterative kernels, which we now discuss.

\paragraph{Choice of target PSF form} It is clear that Airy disks, obscured or unobscured, significantly bias photometric measurements (seventh criterion), while Gaussian functions are not subject to such biases. In Figure~12 of \paptwo, we demonstrated that photometry with {\sc Imcom} is much more accurate than that with {\sc Drizzle}; however, {\sc Imcom} results shown therein were biased towards the ``fainter'' direction. We think we have now successfully identified the cause and the solution to that problem: HSM photometry works less well for images with Airy PSFs; since we have the freedom of choosing PSFs with {\sc Imcom}, it is advisable to adopt Gaussian ones for HSM-like measurements. As for criteria other than control over photometric errors, with the same FWHMs, both versions of Airy disks slightly outperform their Gaussian counterparts in terms of noise control (second, fifth, and sixth criteria); otherwise, there are no significant and consistent differences between PSF forms. See below for how noise control scales with (Gaussian) PSF size.

There is one more point worth discussing about photometry. It is always advisable to perform a post-coaddition flux calibration using objects with known fluxes like injected stars in our tests. Since results with Airy disks are still precise (i.e., have small dispersions), such calibration is expected to largely mitigate or even eliminate the biases. Nevertheless, we note that the seventh to eleventh criteria used in this work are computed for isolated stars; as discussed in Section~4.2 of \paptwo, for realistic images, blending also deteriorates photometric measurements, hence PSFs with significant wings (see Figure~\ref{fig:target_PSFs}) are still disfavored. That said, post-coaddition flux calibration needs to be done for Gaussian target PSFs as well; we leave such effort to future work.

\paragraph{Choice of target PSF size} Comparing Gaussian PSFs with different sizes, we see that wider PSFs lead to many desirable features. When PSFs are narrower by a factor of $20\%$, only input pixels that are close enough to a output pixel can carry significant coaddition weights, which yields poor control over noise. When they are wider by the same factor, the opposite is true; the control over white noise is consistently better by a factor of $\sim 2$ in all bands. More importantly, such benefits do not come at the expense of poor PSF fidelity. With the Cholesky kernel, wider Gaussian PSFs also lead to more precise measurements of injected stars. Most importantly, with the Cholesky kernel, the median ellipticity errors are reduced by factors of $1.94$, $1.35$, $1.83$, $3.48$, and $1.64$ in the Y106, J129, H158, F184, and K213 bands, respectively.

This observation is somewhat counterintuitive, as narrower PSFs (e.g., better seeing conditions for ground-based instruments) are usually better for astronomical observations. One can reconcile these seemingly contradictory facts as follows: Input PSFs determine how a telescope can sample the sky scene, while target output PSFs sets how we prepare images for measurements. See Section~4.1 of \citet{2025arXiv251016110C} for an explanation based the sampling of an underlying optimal ``weight field.'' To address concerns about blending being aggravated by wider PSFs, we note that it is always possible to produce two sets of {\sc Imcom} coadds, one with narrow PSFs for deblending and one with wide (but not too wide) PSFs for measurements. How to balance this with computational costs and storage demands is left for future work.

\subsection{Kernel-specific Settings} \label{ss:kernel-r}

\begin{figure*}
    \centering
    \includegraphics[width=0.9\textwidth]{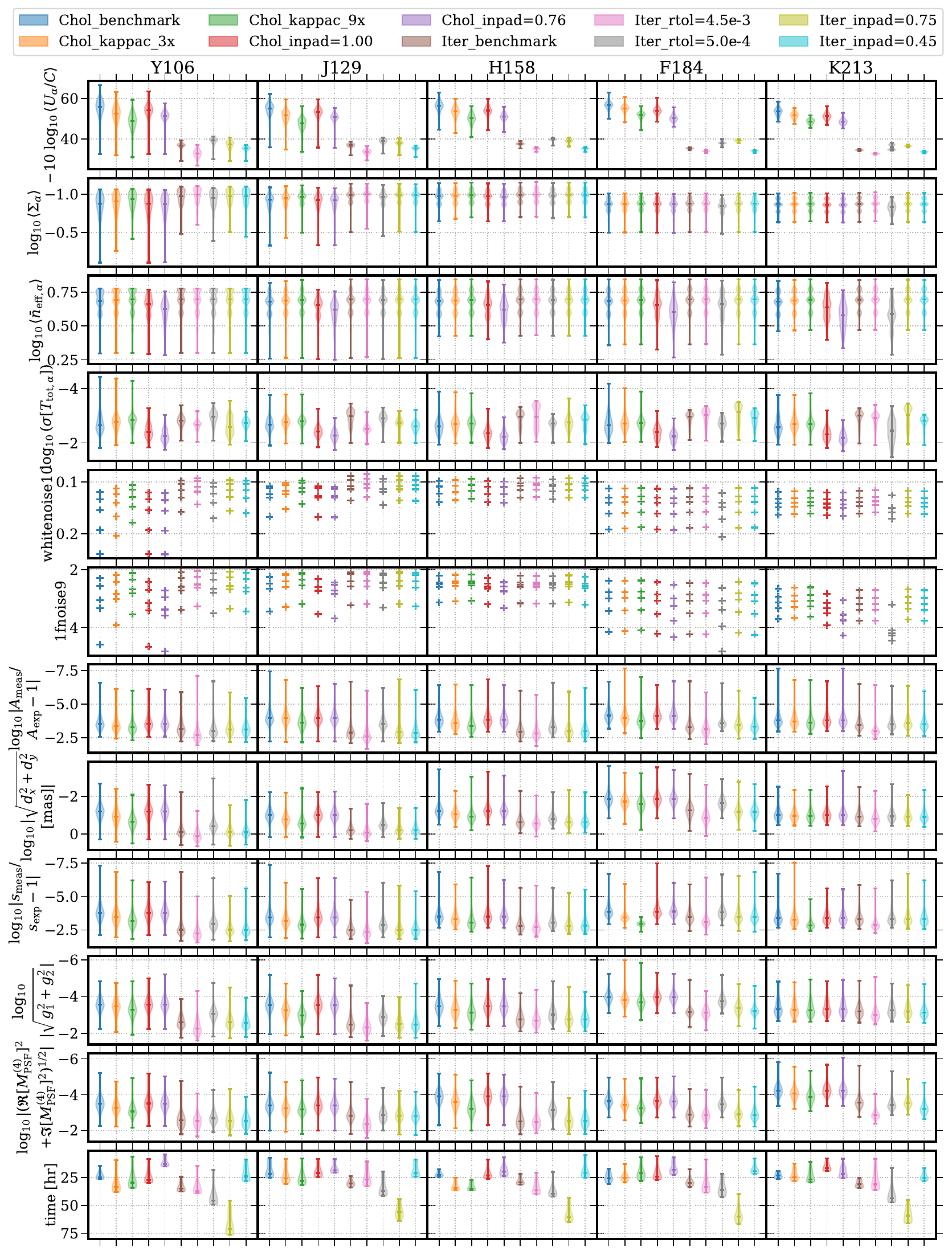}
    \caption{\label{fig:kernel}Similar to Figure~\ref{fig:outpsf}, but comparing different kernel-specific settings.}
\end{figure*}

Figure~\ref{fig:kernel} explores the three linear algebra kernel-specific settings discussed in Section~\ref{ss:kernel-m}.

\paragraph{Lagrange multiplier $\kappa_\alpha$ (for the Cholesky kernel)} By design, the role of the Lagrange multiplier $\kappa_\alpha$ is to balance PSF leakage $U_\alpha$ and noise amplification $\Sigma_\alpha$, both defined in Equation~(\ref{eq:T-AB}). From the first two rows of Figure~\ref{fig:kernel}, we see that both $U_\alpha$ and $\Sigma_\alpha$ are monotonic functions of $\kappa_\alpha$ in all bands, as expected. While a larger $\kappa_\alpha$ value enhances total input weight uniformity (fourth row), it deteriorates all measurements of injected stars. For example, compared to benchmark results, increasing $\kappa_\alpha$ by a factor of $3$ increases median ellipticity errors by factors of $1.27$, $1.93$, $1.57$, $1.40$, and $1.11$ in the five bands, respectively. Therefore, we conclude that larger $\kappa_\alpha$ values worsen the results and are not recommended; however, we caution the readers that they should not be too small either, as otherwise linear systems might be unstable.

\paragraph{Relative tolerance ${\tt rtol}$ (for the iterative kernel)} Intuitively, a lower tolerance is supposed to reduce random errors involved in iterative solutions and thus lead to better measurements. Such expectation is largely met based on what we see in Figure~\ref{fig:kernel}: By reducing ${\tt rtol}$ by a factor of $3$, PSF fidelity, total input weight uniformity, and measurements of injected stars are all enhanced. Specifically, median ellipticity errors are reduced by factors of $2.84$, $4.35$, $1.82$, $3.93$, and $1.11$ in the five bands, respectively; in the F184 band, small-tolerance results of the iterative kernel are almost as good as benchmark results of the Cholesky kernel. While these sound promising, we also see from the last row that decreasing ${\tt rtol}$ significantly increases the time consumption. In the future, if the convergence of the conjugate gradient method for {\sc Imcom} purposes can be made much faster, so that a much smaller ${\tt rtol}$ is affordable, the iterative kernel has the potential of being the better linear algebra strategy thanks to its symmetric selection of input pixels; we leave such algorithmic improvements to future work.

\paragraph{Acceptance radius ${\tt INPAD}$ (for both kernels)} Setting a larger (smaller) acceptance radius amounts to selecting more (less) input pixels for a given postage stamp or output pixel. As argued in Section~\ref{ss:kernel-m}, distant input pixels are likely non-essential, and results shown in Figure~\ref{fig:kernel} allow us to test this intuition. For the Cholesky kernel, although {\sc Imcom} diagnostics noticeably deteriorate when we use smaller acceptance radii, measurements of simulated noise fields and injected stars seem barely affected. Specifically, reducing ${\tt INPAD}$ from $1.24 \,{\rm arcsec}$ to $0.76 \,{\rm arcsec}$ only increases $P_{\rm tot} [{\tt whitenoise10}]$ in the middle mean coverage bin by $1.10\%$, $3.24\%$, $0.83\%$, $0.93\%$, and $1.35\%$ in the five bands, respectively. As for ellipticity errors, the differences are also very small, and surprisingly, the ${\tt INPAD} = 0.76 \,{\rm arcsec}$ results are even slightly better than ${\tt INPAD} = 1.24 \,{\rm arcsec}$ ones in the redder bands. From the last row, we also see that using smaller ${\tt INPAD}$ values indeed reduces time consumption; furthermore, they would make {\sc Imcom} runs with the Cholesky kernel single-core jobs as well. Therefore, reducing the acceptance radius is worth considering.

For the iterative kernel, two alternative acceptance radii are tested, $0.75$ and $0.45$ instead of the benchmark $0.60 \,{\rm arcsec}$. Given the theoretical ${\tt INPAD}^4$ scaling of the complexity of linear system solving, the impact of ${\tt INPAD}$ on time consumption is substantial, as expected. Like for the Cholesky kernel, the measurement results are largely insensitive to the acceptance radius. This supports our above suggestion. Meanwhile, since $0.45 \,{\rm arcsec} \approx 4.09 \,{\tt s\_in}$, it is probably unwise to adopt such a small ${\tt INPAD}$, which is a waste of input information.

\subsection{Experimental Features} \label{ss:exfeat-r}

\begin{figure*}
    \centering
    \includegraphics[width=0.9\textwidth]{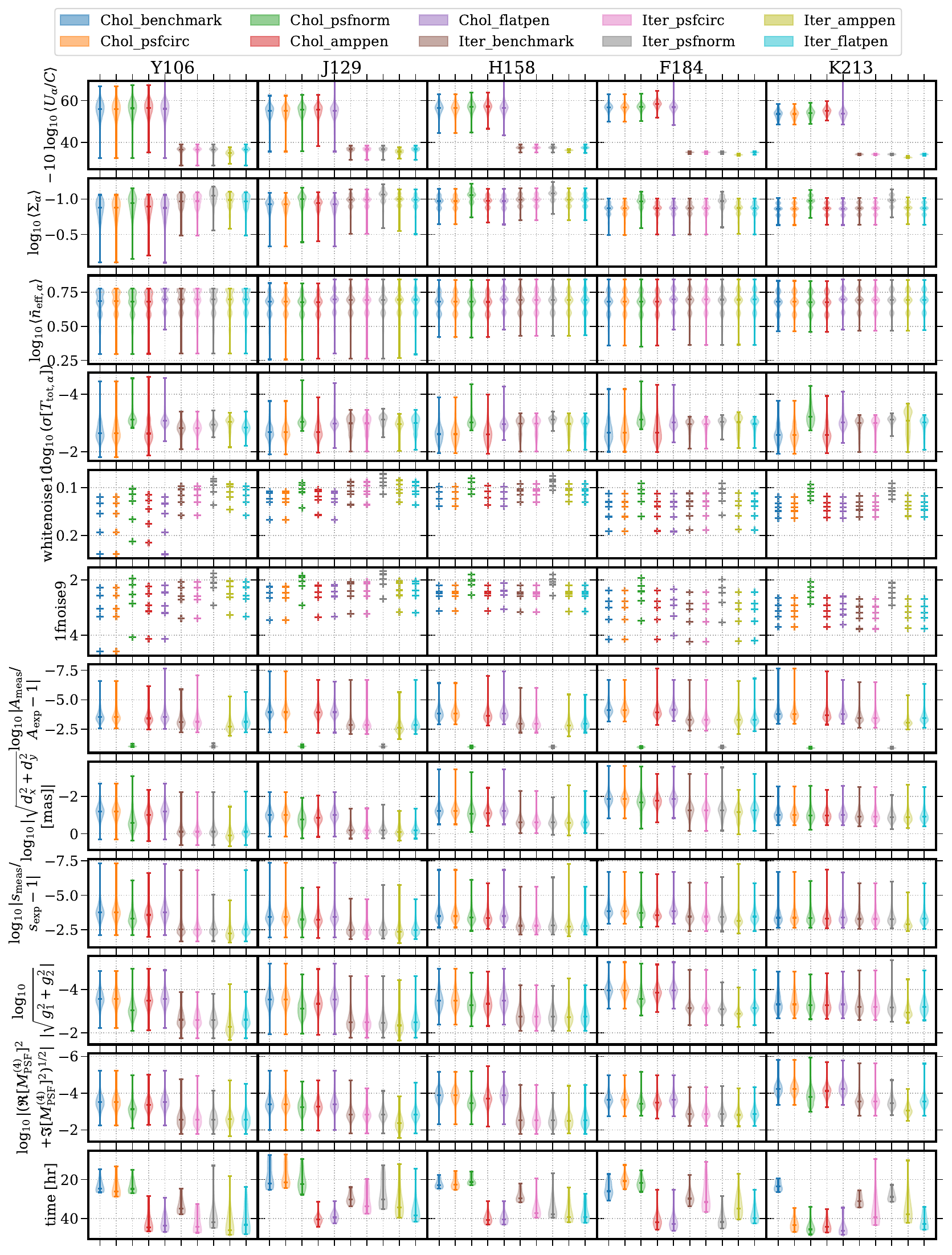}
    \caption{\label{fig:exfeat}Similar to Figure~\ref{fig:outpsf}, but comparing different experimental features.}
\end{figure*}

Figure~\ref{fig:exfeat} tests the four {\sc Imcom} experimental features introduced in Section~\ref{ss:exfeat-m}.

\paragraph{Circular cutouts for PSF arrays ({\tt psfcirc})} For both the Cholesky and iterative kernels, this feature basically has no impact on any of the evaluation criteria. We conclude that the outer regions of PSF arrays are sufficiently close to zero so that whether to make circular cutouts has almost no effect.

\paragraph{Unified normalization of PSF arrays ({\tt psfnorm})} Discrepant normalization of PSF arrays likely results in discrepant contributions from input images. By addressing this issue, the {\tt psfnorm} option enhances both noise control and total input weight uniformity. Nevertheless, it also disrupts the physical meaning of PSFs and corrupts photometric measurements and is probably not worth including.

\paragraph{Penalizing large-amplitude modes ({\tt amppen})} This feature slightly improves noise control, but has no significant impact on other {\sc Imcom} internal diagnostics or measurements of injected stars. Meanwhile, it consistently slows down the program for both linear algebra kernels and in all five bands. Thus we think it is not a useful feature to include in future {\sc Imcom} runs.

\paragraph{Penalizing flat-field differences ({\tt flatpen})} Like {\tt amppen}, this feature causes a similar degree of deceleration. It slightly enhances total input weight uniformity and effective coverage --- which is partially expected, as {\tt flatpen} is supposed to make contributions from different input images less discrepant by design (see Equation~(\ref{eq:flatpen})). However, such enhancements do not translate to more precise measurements, so this option is probably not worth turning on either.

\section{Summary and Discussion} \label{sec:disc}

This paper is a follow-up to our previous papers in this series \citep[][referred to as \papone, \paptwo, and \papthree, respectively]{2024MNRAS.528.2533H, 2024MNRAS.528.6680Y, 2025ApJS..277...55C}, which is dedicated to the application of the {\sc Imcom} algorithm \citep{2011ApJ...741...46R} to the Roman High Latitude Imaging Survey (HLIS). In this work, we have systematically investigated the impact of {\sc Imcom} hyperparameters on the quality of measurement results. We have re-coadded the same $16$ blocks ($1.75 \times 1.75 \,{\rm arcmin}^2$ each) from OpenUniverse2024 simulations \citep[OU24;][]{2025arXiv250105632O} with $13$ different settings with each of the Cholesky and iterative kernels in each of the Y106, J129, H158, F184, and K213 bands. We have compared the results in terms of $12$ objective evaluation criteria, including internal diagnostics of {\sc Imcom}, properties of coadded noise frames, measurements of injected point sources, and time consumption (Section~\ref{ss:eval}). Our major findings can be summarized as follows:
\begin{itemize}
    \item Linear algebra strategy: The Cholesky kernel is the best known strategy for {\sc Imcom} in terms of PSF fidelity, efficiency, and control over errors in measurements of injected objects (Section~\ref{sec:base}). The potential of the iterative kernel can only be realized if a future algorithmic upgrade can make a much smaller tolerance computationally affordable (Section~\ref{ss:kernel-r}).
    \item Target output PSFs (Section~\ref{ss:outpsf-r}): Airy disks, either obscured or unobscured, lead to significant biases in HSM photometric measurements; hence simple Gaussian PSFs are recommended. While some caution is needed, moderately increasing the widths of Gaussian target output PSFs lead to more precise measurements.
    \item Kernel-specific settings (Section~\ref{ss:kernel-r}): For the Cholesky kernel, as long as the linear systems are stable, the Lagrange multiplier $\kappa_\alpha$ in Equation~(\ref{eq:T-AB}) is not worth increasing. The quality of {\sc Imcom} results is not sensitive to the acceptance radius ${\tt INPAD}$, and this can be used to improve computational efficiency.
    \item Experimental features (Section~\ref{ss:exfeat-r}): We have considered two modifications to the PSF arrays in {\sc Imcom} and two penalties to undesirable features, yet they are either inconsequential or detrimental and thus not worth including in future {\sc Imcom} runs.
\end{itemize}

It has been almost two and a half years since we made the to-do list at the end of \paptwo. Before concluding this paper, we would like to briefly review the status of what was proposed there:
\newcounter{SList}
\begin{list}{\arabic{SList}.\ }{\usecounter{SList}}
    \item {\em Computational efficiency}: In \papthree, we have upgraded the software architecture, substituted the bisection search of the optimal Lagrange multiplier $\kappa_\alpha$, and employed several other acceleration measures. In this paper, we have demonstrated the insensitivity of the quality of {\sc Imcom} results to the acceptance radius ${\tt INPAD}$; we may choose smaller values in future {\sc Imcom} runs to save time. Hardware upgrades, e.g., the advent of the Cardinal cluster at the Ohio Supercomputer Center \citep{OhioSupercomputerCenter1987}, have also led to speed-ups. Currently, we expect to spend $\sim 1.1 \times 10^4$ core hours per band per square degree, or $\sim 10^8$ core hours for the combination of ``Medium Tier'' and ``Wider Tier'' \citep{2025arXiv250510574O}; we look forward to making {\sc Imcom} even more efficient through future developments \citep[e.g.,][]{2025arXiv251016110C}.
    \item {\em Extended source injection}: This has been implemented before making the $1.0 \times 1.0 \,{\rm deg}^2$ coadds with OU24 simulated images; see Section~\ref{ss:ou24}. The Shear and Clustering Measurement Working Group of the Roman HLIS Cosmology PIT is actively working on the application of {\sc Metacalibration} \citep{2017arXiv170202600H, 2017ApJ...841...24S} and {\sc Metadetection} \citep{2020ApJ...902..138S} on injected extended sources and simulated science images.
    \item {\em Error propagation}: We have not made significant progress in this area to date, but acknowledge that such investigation is important to better prepare for unanticipated issues.
    \item {\em Laboratory noise fields}: See \citet{2024PASP..136l4506L}.
    \item {\em Poisson noise bias corrections}: See M. Gabe et al. (2025, in preparation).
    \item {\em Chromatic effects}: The impact of chromaticity has been studied by \citet{2025MNRAS.542..608B} in the context of Roman weak lensing. Meanwhile, the propagation of chromatic effects through {\sc Imcom} still needs to be studied, and the corresponding correction schemes remain to be developed.
    \item {\em Deep fields}: As outlined in \paptwo, a natural solution is to coadd subsets of the deep field images and then do a pixel-by-pixel coadd. Grouping input images by epoch, this scheme would allow for studies of secular phenomena like proper motions. Besides, with a ${\cal O}(n^2)$ instead of ${\cal O}(n^3)$ complexity (where $n$ is the number of selected input pixels, which is proportional to the number of contributing images), the iterative kernel introduced in \papthree\ opens up the possibility of coadding all deep field images at once. Nevertheless, further work is needed to address the quality-performance bottleneck of this linear algebra strategy. Furthermore, novel strategies may be developed over the next few years.
    \item {\em Other survey strategies}: The design of the Roman High Latitude Wide Area Survey (HLWAS) has been largely settled \citep{2025arXiv250510574O}. Nevertheless, we hope that {\sc Imcom} will provide successors of the Roman mission with useful lessons.
\end{list}

To conclude this list of to-do items, we note that there are optimal image coaddition algorithms designed for well sampled images \citep[e.g.,][]{2017ApJ...836..187Z, 2017ApJ...836..188Z}. While {\sc Imcom} is designed for undersampled images and has different optimization goals, we hope the strengths of different algorithms can be combined in the future to further advance image processing for weak gravitational lensing cosmology.

\section*{Acknowledgments}

We thank Katarina (Dida) Markovi\v{c}, Arun Kannawadi, Axel Guinot, Sidney Mau, and Federico Berlfein for helping us maintain the {\sc PyImcom} software and Eric Huff and David Weinberg for useful discussions. This paper has undergone internal review in the Roman High Latitude Imaging Survey (HLIS) Cosmology Project Infrastructure Team (PIT). We would like to thank Arun Kannawadi and David Weinberg for helpful comments and feedback during the review process. We thank the anonymous reviewer for insightful comments and Mike Jarvis and Arun Kannawadi for useful discussions during the revision.

K.C., C.H., K.L., E.M., and M.T. received support from the “Maximizing Cosmological Science with the Roman High Latitude Imaging Survey” Roman Project Infrastructure Team (NASA grant 22-ROMAN11-0011). C.H. additionally received support from the David \& Lucile Packard Foundation award 2021-72096. M.Y. and M.T. were also supported by NASA under JPL Contract Task 70-711320, “Maximizing Science Exploitation of Simulated Cosmological Survey Data Across Surveys.”

The detector mask files used in the Roman image simulations are based on data acquired in the Detector Characterization Laboratory (DCL) at the NASA Goddard Space Flight Center. We thank the personnel at the DCL for making the data available for this project.

Computations for this project used the Pitzer cluster at the Ohio Supercomputer Center \citep{Pitzer2018}.

This project made use of the {\sc NumPy} \citep{2020Natur.585..357H}, {\sc Astropy} \citep{2013A&A...558A..33A, 2018AJ....156..123A, 2022ApJ...935..167A}, {\sc SciPy} \citep{2020NatMe..17..261V} and {\sc Numba} \citep{2015llvm.confE...1L} packages.
Most of the figures were made using {\sc Matplotlib} \citep{Hunter:2007}; {\sc SAOImageDS9} \citep{2003ASPC..295..489J} played an important role as a preview tool.
Some of the results in this paper have been derived using the {\sc healpy} and {\sc HEALPix} package \citep{2005ApJ...622..759G, 2019JOSS....4.1298Z}.

\section*{Data Availability}

The codes and configuration files for this project are available in the two GitHub repositories:
\begin{itemize}
\item \url{https://github.com/Roman-HLIS-Cosmology-PIT/pyimcom.git} (introduced in \papthree)
\item \url{https://github.com/Roman-HLIS-Cosmology-PIT/furry-parakeet} (part of \papone\ implementation)
\end{itemize}
This project used {\sc PyImcom} v1.0.3 for simulations, postprocessing, and analysis. It is available on Zenodo under an open-source Creative Commons Attribution license: \dataset[doi: 10.5281/zenodo.17832923]{https://doi.org/10.5281/zenodo.17832923}. C routines in {\sc furry-parakeet} v0.1.1 were used to speed up simulations.

\appendix

\section{Supplemental tables} \label{app:tables}
\renewcommand{\arraystretch}{1.2}

\begin{table*}[]
    \centering
    \caption{\label{tab:whitenoise}Total white noise powers in all $5$ bands $\times 2$ linear algebra kernels $\times 13$ cases. Central values correspond to middle mean coverage bins (see Table~\ref{tab:mean_coverage}), while lower and upper error bars correspond to those of highest and lowest mean coverages, respectively. For visualizations, see the fifth row of Figures~\ref{fig:outpsf}, \ref{fig:kernel}, and \ref{fig:exfeat}.}
    \begin{tabular}{cccccc}
    \hline
        Kernel: Case & Y106 & J129 & H158 & F184 & K213 \\
    \hline
        Chol: benchmark & $0.1542^{+0.0848}_{-0.0346}$ & $0.1234^{+0.0439}_{-0.0150}$ & $0.1088^{+0.0302}_{-0.0108}$ & $0.1397^{+0.0518}_{-0.0269}$ & $0.1412^{+0.0226}_{-0.0219}$ \\
        Chol: airyobsc & $0.1086^{+0.0635}_{-0.0250}$ & $0.0775^{+0.0301}_{-0.0093}$ & $0.0607^{+0.0185}_{-0.0055}$ & $0.0704^{+0.0271}_{-0.0136}$ & $0.0675^{+0.0108}_{-0.0105}$ \\
        Chol: airyunobsc & $0.1252^{+0.0717}_{-0.0286}$ & $0.0964^{+0.0360}_{-0.0116}$ & $0.0820^{+0.0235}_{-0.0079}$ & $0.1029^{+0.0385}_{-0.0198}$ & $0.1035^{+0.0166}_{-0.0162}$ \\
        Chol: gauss{\textunderscore}0.8x & $0.4924^{+0.2952}_{-0.1113}$ & $0.3799^{+0.1631}_{-0.0438}$ & $0.3347^{+0.1151}_{-0.0238}$ & $0.4345^{+0.1762}_{-0.0828}$ & $0.4966^{+0.0703}_{-0.0655}$ \\
        Chol: gauss{\textunderscore}1.2x & $0.0713^{+0.0356}_{-0.0153}$ & $0.0573^{+0.0175}_{-0.0073}$ & $0.0521^{+0.0130}_{-0.0057}$ & $0.0661^{+0.0240}_{-0.0128}$ & $0.0670^{+0.0109}_{-0.0104}$ \\
        Chol: kappac{\textunderscore}3x & $0.1404^{+0.0634}_{-0.0282}$ & $0.1170^{+0.0358}_{-0.0149}$ & $0.1073^{+0.0279}_{-0.0114}$ & $0.1391^{+0.0507}_{-0.0267}$ & $0.1400^{+0.0223}_{-0.0218}$ \\
        Chol: kappac{\textunderscore}9x & $0.1293^{+0.0496}_{-0.0234}$ & $0.1117^{+0.0308}_{-0.0148}$ & $0.1060^{+0.0264}_{-0.0118}$ & $0.1383^{+0.0497}_{-0.0264}$ & $0.1388^{+0.0220}_{-0.0216}$ \\
        Chol: inpad=1.00 & $0.1547^{+0.0846}_{-0.0345}$ & $0.1267^{+0.0410}_{-0.0179}$ & $0.1091^{+0.0302}_{-0.0108}$ & $0.1401^{+0.0519}_{-0.0270}$ & $0.1419^{+0.0225}_{-0.0220}$ \\
        Chol: inpad=0.76 & $0.1559^{+0.0842}_{-0.0343}$ & $0.1274^{+0.0409}_{-0.0135}$ & $0.1097^{+0.0302}_{-0.0108}$ & $0.1410^{+0.0519}_{-0.0270}$ & $0.1431^{+0.0225}_{-0.0223}$ \\
        Chol: psfcirc & $0.1542^{+0.0848}_{-0.0346}$ & $0.1234^{+0.0439}_{-0.0150}$ & $0.1088^{+0.0302}_{-0.0108}$ & $0.1397^{+0.0518}_{-0.0269}$ & $0.1412^{+0.0226}_{-0.0219}$ \\
        Chol: psfnorm & $0.1280^{+0.0839}_{-0.0254}$ & $0.1031^{+0.0393}_{-0.0129}$ & $0.0894^{+0.0240}_{-0.0085}$ & $0.1126^{+0.0476}_{-0.0210}$ & $0.1093^{+0.0178}_{-0.0159}$ \\
        Chol: amppen & $0.1452^{+0.0700}_{-0.0303}$ & $0.1191^{+0.0381}_{-0.0149}$ & $0.1078^{+0.0285}_{-0.0112}$ & $0.1394^{+0.0511}_{-0.0268}$ & $0.1404^{+0.0224}_{-0.0218}$ \\
        Chol: flatpen & $0.1544^{+0.0849}_{-0.0347}$ & $0.1235^{+0.0441}_{-0.0150}$ & $0.1088^{+0.0303}_{-0.0108}$ & $0.1398^{+0.0519}_{-0.0270}$ & $0.1413^{+0.0226}_{-0.0220}$ \\
    \hline
        Iter: benchmark & $0.1161^{+0.0422}_{-0.0191}$ & $0.1077^{+0.0282}_{-0.0195}$ & $0.1066^{+0.0238}_{-0.0140}$ & $0.1386^{+0.0498}_{-0.0265}$ & $0.1392^{+0.0221}_{-0.0219}$ \\
        Iter: airyobsc & $0.0791^{+0.0285}_{-0.0131}$ & $0.0667^{+0.0175}_{-0.0119}$ & $0.0586^{+0.0131}_{-0.0075}$ & $0.0691^{+0.0249}_{-0.0132}$ & $0.0662^{+0.0106}_{-0.0104}$ \\
        Iter: airyunobsc & $0.0925^{+0.0336}_{-0.0153}$ & $0.0832^{+0.0218}_{-0.0150}$ & $0.0798^{+0.0178}_{-0.0104}$ & $0.1020^{+0.0367}_{-0.0195}$ & $0.1025^{+0.0163}_{-0.0161}$ \\
        Iter: gauss{\textunderscore}0.8x & $0.3677^{+0.1502}_{-0.0643}$ & $0.3324^{+0.0976}_{-0.0529}$ & $0.3148^{+0.0804}_{-0.0376}$ & $0.3989^{+0.1578}_{-0.0749}$ & $0.4372^{+0.0644}_{-0.0663}$ \\
        Iter: gauss{\textunderscore}1.2x & $0.0535^{+0.0167}_{-0.0084}$ & $0.0514^{+0.0123}_{-0.0096}$ & $0.0520^{+0.0115}_{-0.0068}$ & $0.0658^{+0.0237}_{-0.0127}$ & $0.0664^{+0.0107}_{-0.0104}$ \\
        Iter: rtol=4.5e-3 & $0.1089^{+0.0346}_{-0.0166}$ & $0.1044^{+0.0252}_{-0.0199}$ & $0.1059^{+0.0237}_{-0.0139}$ & $0.1378^{+0.0495}_{-0.0264}$ & $0.1374^{+0.0219}_{-0.0216}$ \\
        Iter: rtol=5.0e-4 & $0.1229^{+0.0475}_{-0.0215}$ & $0.1133^{+0.0313}_{-0.0192}$ & $0.1084^{+0.0255}_{-0.0138}$ & $0.1495^{+0.0569}_{-0.0283}$ & $0.1510^{+0.0205}_{-0.0254}$ \\
        Iter: inpad=0.75 & $0.1157^{+0.0404}_{-0.0191}$ & $0.1079^{+0.0278}_{-0.0195}$ & $0.1050^{+0.0254}_{-0.0123}$ & $0.1384^{+0.0498}_{-0.0265}$ & $0.1391^{+0.0221}_{-0.0220}$ \\
        Iter: inpad=0.45 & $0.1161^{+0.0438}_{-0.0193}$ & $0.1065^{+0.0299}_{-0.0182}$ & $0.1066^{+0.0239}_{-0.0140}$ & $0.1386^{+0.0496}_{-0.0266}$ & $0.1403^{+0.0223}_{-0.0220}$ \\
        Iter: psfcirc & $0.1161^{+0.0422}_{-0.0191}$ & $0.1077^{+0.0282}_{-0.0195}$ & $0.1066^{+0.0238}_{-0.0140}$ & $0.1386^{+0.0498}_{-0.0265}$ & $0.1392^{+0.0221}_{-0.0219}$ \\
        Iter: psfnorm & $0.0946^{+0.0419}_{-0.0124}$ & $0.0885^{+0.0252}_{-0.0168}$ & $0.0882^{+0.0176}_{-0.0120}$ & $0.1116^{+0.0457}_{-0.0206}$ & $0.1073^{+0.0173}_{-0.0158}$ \\
        Iter: amppen & $0.1095^{+0.0364}_{-0.0165}$ & $0.1048^{+0.0253}_{-0.0200}$ & $0.1061^{+0.0237}_{-0.0139}$ & $0.1381^{+0.0496}_{-0.0264}$ & $0.1377^{+0.0218}_{-0.0218}$ \\
        Iter: flatpen & $0.1161^{+0.0422}_{-0.0191}$ & $0.1077^{+0.0282}_{-0.0195}$ & $0.1066^{+0.0238}_{-0.0140}$ & $0.1386^{+0.0499}_{-0.0265}$ & $0.1392^{+0.0221}_{-0.0219}$ \\
    \hline
    \end{tabular}
\end{table*}

\begin{table*}[]
    \centering
    \caption{\label{tab:ellipticity}Logarithmic ellipticities of $957$ injected stars in all $5$ bands $\times 2$ linear algebra kernels $\times 13$ cases. Central values correspond to medians, while lower and upper error bars correspond to the $16$th and $84$th percentiles, respectively. For distributions, see the tenth (i.e., third-to-last) row of Figures~\ref{fig:outpsf}, \ref{fig:kernel}, and \ref{fig:exfeat}.}
    \begin{tabular}{cccccc}
    \hline
        Kernel: Case & Y106 & J129 & H158 & F184 & K213 \\
    \hline
        Chol: benchmark & $-3.5636^{+0.3349}_{-0.3595}$ & $-3.5357^{+0.3337}_{-0.3481}$ & $-3.4792^{+0.3440}_{-0.4199}$ & $-3.9646^{+0.2805}_{-0.3137}$ & $-3.3155^{+0.2985}_{-0.4796}$ \\
        Chol: airyobsc & $-3.6104^{+0.3298}_{-0.3628}$ & $-3.6373^{+0.3354}_{-0.3509}$ & $-3.6049^{+0.3452}_{-0.4237}$ & $-4.0716^{+0.2782}_{-0.3025}$ & $-3.3669^{+0.3092}_{-0.4776}$ \\
        Chol: airyunobsc & $-3.5820^{+0.3300}_{-0.3626}$ & $-3.5983^{+0.3356}_{-0.3448}$ & $-3.5315^{+0.3488}_{-0.4123}$ & $-3.9982^{+0.2797}_{-0.3077}$ & $-3.3286^{+0.3034}_{-0.4819}$ \\
        Chol: gauss{\textunderscore}0.8x & $-3.1699^{+0.3346}_{-0.3770}$ & $-3.3469^{+0.3304}_{-0.3424}$ & $-3.2136^{+0.3462}_{-0.4125}$ & $-3.4008^{+0.2766}_{-0.3056}$ & $-3.0840^{+0.2844}_{-0.4695}$ \\
        Chol: gauss{\textunderscore}1.2x & $-3.8511^{+0.3553}_{-0.3369}$ & $-3.6658^{+0.3566}_{-0.3440}$ & $-3.7414^{+0.3218}_{-0.4428}$ & $-4.5059^{+0.2972}_{-0.3544}$ & $-3.5303^{+0.3031}_{-0.4896}$ \\
        Chol: kappac{\textunderscore}3x & $-3.4609^{+0.3625}_{-0.3683}$ & $-3.2494^{+0.3471}_{-0.3230}$ & $-3.2833^{+0.3386}_{-0.4218}$ & $-3.8198^{+0.2850}_{-0.3112}$ & $-3.2705^{+0.2917}_{-0.4661}$ \\
        Chol: kappac{\textunderscore}9x & $-3.2771^{+0.3590}_{-0.3877}$ & $-2.9827^{+0.3487}_{-0.3479}$ & $-3.1200^{+0.3316}_{-0.4063}$ & $-3.6949^{+0.3057}_{-0.3059}$ & $-3.2552^{+0.2969}_{-0.4632}$ \\
        Chol: inpad=1.00 & $-3.5630^{+0.3336}_{-0.3572}$ & $-3.5353^{+0.3338}_{-0.3466}$ & $-3.4793^{+0.3438}_{-0.4207}$ & $-3.9657^{+0.2811}_{-0.3133}$ & $-3.3162^{+0.2986}_{-0.4812}$ \\
        Chol: inpad=0.76 & $-3.5611^{+0.3292}_{-0.3603}$ & $-3.5355^{+0.3342}_{-0.3460}$ & $-3.4799^{+0.3457}_{-0.4180}$ & $-3.9676^{+0.2812}_{-0.3080}$ & $-3.3162^{+0.2988}_{-0.4794}$ \\
        Chol: psfcirc & $-3.5636^{+0.3349}_{-0.3595}$ & $-3.5357^{+0.3337}_{-0.3481}$ & $-3.4792^{+0.3440}_{-0.4199}$ & $-3.9646^{+0.2805}_{-0.3137}$ & $-3.3155^{+0.2985}_{-0.4796}$ \\
        Chol: psfnorm & $-3.0495^{+0.3505}_{-0.4555}$ & $-3.1168^{+0.3846}_{-0.4487}$ & $-3.2655^{+0.2927}_{-0.3213}$ & $-3.5673^{+0.2208}_{-0.3633}$ & $-3.2843^{+0.2878}_{-0.3864}$ \\
        Chol: amppen & $-3.4886^{+0.3489}_{-0.3792}$ & $-3.3415^{+0.3383}_{-0.3235}$ & $-3.3420^{+0.3399}_{-0.4178}$ & $-3.8610^{+0.2832}_{-0.3124}$ & $-3.2780^{+0.2951}_{-0.4663}$ \\
        Chol: flatpen & $-3.5634^{+0.3348}_{-0.3605}$ & $-3.5356^{+0.3338}_{-0.3480}$ & $-3.4792^{+0.3440}_{-0.4199}$ & $-3.9638^{+0.2798}_{-0.3142}$ & $-3.3155^{+0.2985}_{-0.4796}$ \\
    \hline
        Iter: benchmark & $-2.6100^{+0.2537}_{-0.3691}$ & $-2.4894^{+0.2608}_{-0.4002}$ & $-2.7608^{+0.2935}_{-0.3912}$ & $-3.1527^{+0.1786}_{-0.2265}$ & $-3.1980^{+0.2688}_{-0.3481}$ \\
        Iter: airyobsc & $-2.5689^{+0.2688}_{-0.3894}$ & $-2.5242^{+0.2855}_{-0.4116}$ & $-2.8297^{+0.2910}_{-0.4005}$ & $-3.1097^{+0.1512}_{-0.2051}$ & $-3.1651^{+0.2519}_{-0.2989}$ \\
        Iter: airyunobsc & $-2.5896^{+0.2601}_{-0.3751}$ & $-2.4821^{+0.2618}_{-0.4124}$ & $-2.7604^{+0.2968}_{-0.3970}$ & $-3.1253^{+0.1716}_{-0.2086}$ & $-3.1843^{+0.2783}_{-0.3250}$ \\
        Iter: gauss{\textunderscore}0.8x & $-2.6951^{+0.2864}_{-0.3300}$ & $-2.5938^{+0.2817}_{-0.3118}$ & $-2.7291^{+0.3221}_{-0.4348}$ & $-3.0847^{+0.2556}_{-0.3153}$ & $-2.9745^{+0.2841}_{-0.3586}$ \\
        Iter: gauss{\textunderscore}1.2x & $-2.5146^{+0.2926}_{-0.4597}$ & $-2.6542^{+0.3059}_{-0.3612}$ & $-3.0827^{+0.2646}_{-0.3158}$ & $-3.3055^{+0.2404}_{-0.3090}$ & $-3.1750^{+0.2099}_{-0.2937}$ \\
        Iter: rtol=4.5e-3 & $-2.2584^{+0.3078}_{-0.4538}$ & $-2.3173^{+0.2781}_{-0.3641}$ & $-2.6985^{+0.2804}_{-0.3505}$ & $-3.1207^{+0.2605}_{-0.3203}$ & $-2.9952^{+0.2333}_{-0.3255}$ \\
        Iter: rtol=5.0e-4 & $-3.0641^{+0.2961}_{-0.3151}$ & $-2.8876^{+0.2912}_{-0.3425}$ & $-3.0202^{+0.3093}_{-0.4111}$ & $-3.7468^{+0.2368}_{-0.3194}$ & $-3.2432^{+0.2930}_{-0.4120}$ \\
        Iter: inpad=0.75 & $-2.6133^{+0.2615}_{-0.3719}$ & $-2.4994^{+0.2552}_{-0.4079}$ & $-2.7643^{+0.2921}_{-0.3858}$ & $-3.3690^{+0.2861}_{-0.3168}$ & $-3.2018^{+0.2689}_{-0.3883}$ \\
        Iter: inpad=0.45 & $-2.5771^{+0.2521}_{-0.3839}$ & $-2.4871^{+0.2556}_{-0.4096}$ & $-2.7668^{+0.3036}_{-0.3739}$ & $-3.2076^{+0.1897}_{-0.2426}$ & $-3.1306^{+0.2276}_{-0.3127}$ \\
        Iter: psfcirc & $-2.6057^{+0.2474}_{-0.3757}$ & $-2.4904^{+0.2610}_{-0.3976}$ & $-2.7608^{+0.2931}_{-0.3912}$ & $-3.1527^{+0.1791}_{-0.2265}$ & $-3.1980^{+0.2688}_{-0.3481}$ \\
        Iter: psfnorm & $-2.5960^{+0.2383}_{-0.3613}$ & $-2.4732^{+0.2500}_{-0.4035}$ & $-2.7530^{+0.2930}_{-0.3957}$ & $-3.0869^{+0.1601}_{-0.2342}$ & $-3.1720^{+0.2546}_{-0.3583}$ \\
        Iter: amppen & $-2.2740^{+0.2949}_{-0.4665}$ & $-2.3488^{+0.3016}_{-0.4146}$ & $-2.7188^{+0.2837}_{-0.3687}$ & $-2.8882^{+0.1306}_{-0.1905}$ & $-2.9453^{+0.1746}_{-0.3082}$ \\
        Iter: flatpen & $-2.6100^{+0.2529}_{-0.3695}$ & $-2.4931^{+0.2654}_{-0.3910}$ & $-2.7608^{+0.2959}_{-0.3914}$ & $-3.1527^{+0.1757}_{-0.2270}$ & $-3.2002^{+0.2728}_{-0.3421}$ \\
    \hline
    \end{tabular}
\end{table*}

This appendix contains Tables~\ref{tab:whitenoise} and \ref{tab:ellipticity} to supplement results presented in Sections~\ref{sec:base} and \ref{sec:var}. See those two sections for further explanations and discussions.

\bibliography{main4}{}
\bibliographystyle{aasjournalv7}

\end{document}